\documentclass[11pt,a4paper,abstract=on]{scrartcl}

\usepackage[utf8]{inputenc}
\usepackage[english]{babel}

\usepackage[paper=a4paper,left=25mm,right=25mm,top=25mm,bottom=30mm]{geometry}
\usepackage{amsmath,amssymb,amsthm,mathrsfs,bbm,amsfonts,bm}
\usepackage{color,graphicx,cases,caption,enumerate}
\usepackage{float,multicol,authblk}
\usepackage[dvipsnames]{xcolor}
\usepackage{colortbl}
\usepackage{booktabs}
\usepackage{multirow}
\usepackage[title]{appendix}
\usepackage{subcaption}
\usepackage{booktabs}
\usepackage{bbm}
\usepackage{array}
\usepackage{natbib}
\usepackage[breaklinks, colorlinks=true, citecolor=black, urlcolor=black, linkcolor=black]{hyperref}

\usepackage{siunitx}
\usepackage{comment}
\usepackage{changepage}

\hypersetup{urlcolor=cyan}
\title{Deep learning for post-processing global probabilistic forecasts on sub-seasonal \\ time scales}

\author[1]{Nina Horat}
\author[1,2]{Sebastian Lerch}
\affil[1]{Karlsruhe Institute of Technology}
\affil[2]{Heidelberg Institute for Theoretical Studies}

\date{\today}

\begin{document}
	
	\maketitle
	
	\begin{abstract}
		Sub-seasonal weather forecasts are becoming increasingly important for a range of socio-economic activities. However, the predictive ability of physical weather models is very limited on these time scales.
		We propose several post-processing methods based on convolutional neural networks to improve sub-seasonal forecasts by correcting systematic errors of numerical weather prediction models. 
		Our post-processing models operate directly on spatial input fields and are therefore able to retain spatial relationships and to generate spatially homogeneous predictions. 
		They produce global probabilistic tercile forecasts for biweekly aggregates of temperature and precipitation for weeks 3-4 and 5-6. 
		In a case study based on a public forecasting challenge organized by the World Meteorological Organization, our post-processing models outperform recalibrated forecasts from the European Centre for Medium-Range Weather Forecasts (ECMWF), and achieve improvements over climatological forecasts for all considered variables and lead times.
		We compare several model architectures and training modes and demonstrate that all approaches lead to skillful and well-calibrated probabilistic forecasts.
		The good calibration of the post-processed forecasts emphasizes that our post-processing models
		reliably quantify the forecast uncertainty based on deterministic input information in form of the ECMWF ensemble mean forecast fields only.
		
	\end{abstract}
	
	\section{Introduction}
	Planning and decision making in public health, agriculture, energy supply, water resource management, and other weather-dependent sectors often happens  weeks to months in advance \citep{White2017, White2022applications}. However, due to the lack of forecast skill of physics-based numerical weather prediction (NWP) models on the sub-seasonal to seasonal (S2S) time scale, stake-holders so far mostly rely on either short- to medium-range weather forecasts or on seasonal outlooks \citep{White2022applications}. Therefore, more research on improving S2S forecasts is urgently needed. One important and widely-researched approach is to better identify and exploit windows of opportunity, i.e., times with higher predictability associated with modes of intra-seasonal and seasonal variability, for example MJO and ENSO \citep[e.g.,][]{Mariotti2020, robertson2020, mayer_subseasonal_2021}. On the other hand, the use of post-processing models to correct systematic errors of NWP forecasts has become a standard practice in research and operations in short- to medium-range weather forecasting \citep{vannitsem_statistical_2021}. However, there is a remarkable lack of post-processing approaches for S2S forecasting \citep{robertson2020}, even though it seems reasonable to expect improvements also on these time scales \citep{Merryfield2020s2s}.
	
	Over the past years, the use of modern machine learning (ML) techniques such as random forests \citep{TaillardatEtAl2016}, gradient boosting \citep{MessnerEtAl2017} or neural networks \citep{rasp_neural_2018}  has become a key focal point of research activities in post-processing \citep{haupt2021, vannitsem_statistical_2021}. In many applications to medium-range forecasting, in particular neural network (NN)-based methods have superseded traditional statistical approaches and show substantial improvements in predictive performance. The superior performance of the ML models can mainly be attributed to their ability to better incorporate arbitrary input predictors and to more flexibly model nonlinear relationships. 
	
	One major challenge in statistical post-processing is to retain spatial and temporal relationships in the post-processed forecasts, as well as relationships between variables \citep{vannitsem_statistical_2021}. In particular, the few examples of S2S post-processing studies that we are aware of tend to separately operate on single grid-cells only, and thus are neither able to exploit spatial information in the raw forecasts, nor produce spatially homogeneous forecasts \citep[e.g.,][]{Vigaud2017MMEprecip, Vigaud2019MMEtemperature, mouatadid_learned_2021,  zhang_improving_2023}. For short- to medium-range forecasts, convolutional neural network (CNN)-based architectures and model components have been used for a variety of post-processing applications \citep{DaiHemri2021, gronquist2021, Veldkamp2021wind, Chapman2022, LerchPolsterer2022, Li2022precip, PoET2023, Hu2023precip}. CNN architectures are designed for image-like data and can operate on spatial data directly. Therefore, they enable the generation of spatially homogeneous predictions and have the ability to extract and learn spatial error structures. 
	
	As noted above, research on S2S post-processing is scarce. One of the rare exceptions is the study of \citet{scheuererBasisfunc} who propose a CNN architecture that estimates coefficient values for a set of basis functions to create spatial forecasts for precipitation over California on S2S time scales. 
	In 2021, the World Meteorological Organization (WMO) coordinated a challenge to assess and promote the use of artificial intelligence (AI) for improving S2S forecasts \citep{s2saiChallenge}. The data provided within the framework of the challenge mainly consists of ML-ready (re-)forecasts from the European Centre for Medium-Range Weather Forecasts (ECMWF). 
	The setup of the challenge and the corresponding data provide a convenient starting point for the development of ML-based post-processing methods for S2S forecasts, which produce probabilistic, spatially coherent and calibrated forecasts. 
	
	We propose four CNN-based post-processing methods for global temperature and precipitation forecasts with a lead time of two and four weeks. Our models utilize spatial forecast fields from the ECMWF predictions of several weather variables as input, and provide global probabilistic predictions for terciles (below normal, near normal, and above normal conditions) as commonly done in S2S forecasting \citep[e.g.,][]{Vigaud2017MMEprecip, Vigaud2019MMEtemperature, Mariotti2020, robertson2020, White2022applications}. We train our post-processing models either directly on the full global input fields, or on smaller quadratic subdomains. To the best of our knowledge, the proposed models are the first to solely operate on global spatial inputs, without using any grid cell-specific model components. Using the setup of the WMO-organized S2S AI Challenge as case studies, we compare our CNN models to climatological and ECMWF reference forecasts, and discuss the results within the context of the challenge submissions, which were largely based on grid cell-specific approaches \citep{s2saiChallenge}.
	
	The remainder of this paper is organized as follows. Section \ref{data} introduces the S2S AI Challenge data and setup, and Section \ref{benchmarks} discusses the benchmark methods. Our CNN-based post-processing models are described in Section \ref{sec:CNN}. The main results are presented in Section \ref{sec:results}, and the paper concludes with a discussion in Section \ref{sec:conclusions}.
	Python code with implementations of all methods is available at \url{https://github.com/HoratN/pp-s2s}. 

	\section{Data}\label{data}
	
	The data used for this study stems from the \textit{WMO Prize Challenge to Improve Sub-Seasonal to Seasonal Predictions Using Artificial Intelligence}\footnote{\url{https://s2s-ai-challenge.github.io/}} that took place from June to November 2021 \citep{s2saiChallenge} and which we will refer to as \textit{S2S AI Challenge} in the following. This challenge was coordinated by WMO to foster the use of AI for post-processing S2S forecasts, and to promote the S2S database \citep{s2sProject} for training AI methods. The data set consists of global forecasts and reforecasts for precipitation and temperature from ECMWF. Weekly ensemble reforecasts  with 11 members are provided for the years 2000 to 2019, which were to be used as training data, and forecasts with 51 members for the year 2020, which served as a test data.  The process of training post-processing models on re-forecasts and applying them to forecasts poses methodological challenges, for example because of the different number of ensemble members, but is a common approach in operational weather forecasting \citep[e.g.,][]{PoET2023,DemaeyerEtAl2023}. 
	
	All predictions come at a spatial resolution of 1.5$^\circ$ and a temporal resolution of 2 weeks. The biweekly aggregated (re-)forecasts have a lead time of 14 and 28 days, respectively, and are computed by averaging daily temperature over days 15 to 28 (for week 3-4), and days 29 to 42 (for week 5-6) as described in \citep{s2saiChallenge}. Biweekly precipitation (re-)forecasts are obtained by accumulating daily precipitation rates over the same periods. The challenge team provided verifying gridded observations over land from the NOAA Climate Prediction Center~(CPC) for precipitation and temperature. 
	The observational and \mbox{(re-)}forecast data is available through the S2S AI Challenge repository\footnote{\url{https://renkulab.io/gitlab/aaron.spring/s2s-ai-challenge-template/-/tree/master/data}} and a dedicated CliMetLab plugin that provides forecast data for additional variables, and from other forecasting centers\footnote{\url{https://github.com/ecmwf-lab/climetlab-s2s-ai-challenge}}.
	
	The main prediction task within the framework of the S2S AI Challenge was to provide categorical probability forecasts for ``below normal'', ``near normal'', and ``above normal''  values of the target variables. To define these categories, the challenge team computed category edges based on the empirical 1/3~- and 2/3~-quantiles of the observations from the training period (years 2000 to 2019). 
	These edges were computed separately for each grid cell and week of the year to account for spatial and seasonal differences.

	\section{Benchmark methods}\label{benchmarks}
	
	\subsection{Climatological forecasts}
	
	An exceedingly simple, yet often competitive benchmark in sub-seasonal to seasonal forecasting is the climatological forecast. In the setting described above, the probabilistic climatological forecasts simply assigns a probability of 1/3 to each of the three categories. Given the computation of the category edges based on historical data, the climatological forecast can generally be expected to be calibrated.
	
	\subsection{ECMWF baseline and corrections}
	
	The S2S AI Challenge also provided a more advanced benchmark method for the year 2020 based on ECMWF ensemble forecasts, henceforth called ECMWF baseline. 
	Categorical probability forecasts are obtained by computing the empirical frequencies of the 51-member ensemble after binning the members into three categories.
	To enforce calibrated forecasts, the binning is done with respect to the category edges computed from the re-forecasts instead of using the respective category edges based on historical observations. 
	For very dry regions (accumulated precipitation over two weeks $<$ \SI{0.01}{mm/m^2}), a climatological forecast is issued for precipitation since the tercile edges are too close together, preventing a robust computation. 
	
\begin{figure}
		\centering
		\includegraphics[width=0.9\textwidth]{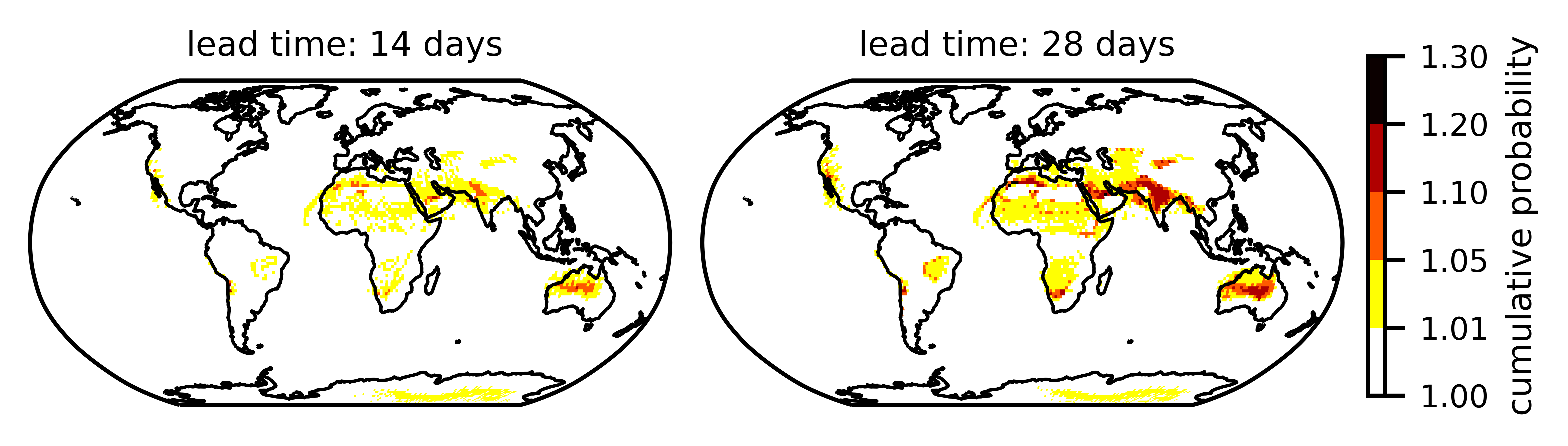}
		\caption{Sum of the forecast probabilities for the three tercile categories as provided by the ECMWF baseline within the framework of the S2S AI Challenge. Shown is the average  over all forecasts from 2020.}
		\label{fig:baseline}
	\end{figure}

	The ECMWF baseline provided within the framework of the S2S AI Challenge comes with two limitations. First, it is only available for the test data set (the year 2020). Second, a closer inspection of the precipitation forecasts revealed that the probabilities over the three categories sometimes do not sum up to 1 due to an erroneous treatment of missing values in the ensemble members.
	This problematic behavior becomes apparent in Figure \ref{fig:baseline}, which shows the sum of the predicted probabilities of the three categories for all grid points. Despite being averaged over the entire year 2020, substantial deviations from 1 can be observed for a large fraction of grid points.
	
	Based on the Fortran code used to create the ECMWF baseline (Frederic Vitart, personal communication), we implemented our own baseline method following the same idea, but accounting for members with missing values by computing the tercile probabilities based on the number of non-missing members only. For very dry regions, we follow the construction of the ECMWF baseline and issue a climatological forecast for precipitation. We refer to this benchmark as corrected ECMWF baseline.

	\section{Convolutional Neural Network architectures}\label{sec:CNN}
	
	We propose four different post-processing methods for gridded data that are based on CNN architectures. 
	All architectures follow an encoder-decoder structure. The encoder part detects structures in the global gridded input data and extracts lower-dimensional latent features from the image-like input. The decoder part can then make use of the encoded information to create a spatial prediction, providing a probabilistic forecast for tercile categories at every land grid cell globally. 
	
	The encoder part of our CNNs uses alternating layers of convolutional and pooling operations. Convolutional layers allow the model to learn and detect structures in image-like input data. Instead of learning separate weights for each pixel in the image (as a dense neural network would do), weights are shared among pixels by using so-called filters. These filters are three-dimensional weight tensors learned during training. The third dimension corresponds to the number of input maps; the number of the filters corresponds to the number of different structures the model can detect. To detect the structures, the filter tensor slides over the image and a dot product between a part of the image and the filter is computed. Between the convolutional layers, pooling layers are used that aggregate grid points to amplify the signal found by the convolutional layers. With every convolutional and pooling layer pair, the image resolution decreases and more and more complex structures are learnt from the input image. While this feature extraction part, the encoder, varies only slightly for the four different methods proposed here, the decoder part of the four post-processing architectures differs significantly. 
	
\begin{figure}
		\centering
		\includegraphics[width=\textwidth,trim={1cm 5cm 1.5cm 6cm},clip]{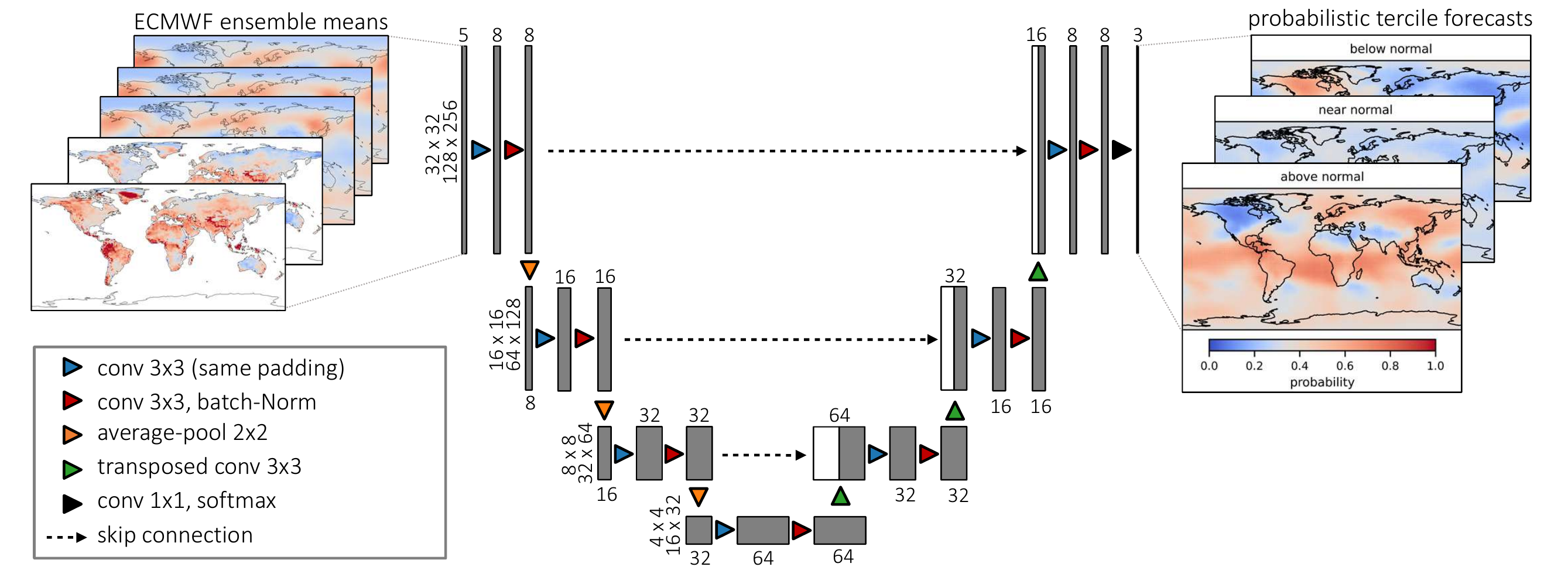}
		\caption{The UNet architecture used for temperature (for precipitation, one additional predictor is used, resulting in six inputs). The number of global maps (channels) is denoted on top of each box, the spatial extent of these feature maps is given on the left side of each box for the patch-wise training (left number) and the global training (right number).}
		\label{fig:schematicsUnet}
	\end{figure}

	In the following we briefly summarize the key ideas behind the four post-processing models. The first architecture that we propose is based on the concept of a UNet, a specific type of a CNN architecture that is used for image segmentation and was first proposed in medical applications \citep{unet,imageSegmetationReview}. In atmospheric sciences,  it  is also gaining popularity in particular for detection, nowcasting and forecasting \citep[e.g.,][]{ayzel2020,  lagerquist2021, Chapman2022, quinting2022}. The UNet architecture has a symmetric encoder-decoder architecture (see Figure \ref{fig:schematicsUnet} for a schematic illustration of the model architecture) and is specifically tailored to image-to-image tasks. 
	In contrast to traditional CNN model architectures for image recognition tasks, UNets do not classify an image as a whole but instead assign a class probability to every pixel. 
	Therefore, UNets can be used to directly obtain probabilities for tercile categories jointly at all grid cells. 
	In a UNet, the up-scaling from the learned low-dimensional representation to the full image size (i.e., the decoding) is done using so-called up-sampling layers, for which we use transposed convolutions. The output of the up-sampling layer is concatenated with the feature maps of the same size from the feature-extraction (encoder) part of the model (as depicted in Figure \ref{fig:schematicsUnet}). These concatenations are called skip connections and help to reconstruct the fine details from the high-resolution input.
	
\begin{figure}
		\centering
		\includegraphics[width=\textwidth,trim={1cm 2.5cm 1.5cm 5cm},clip]{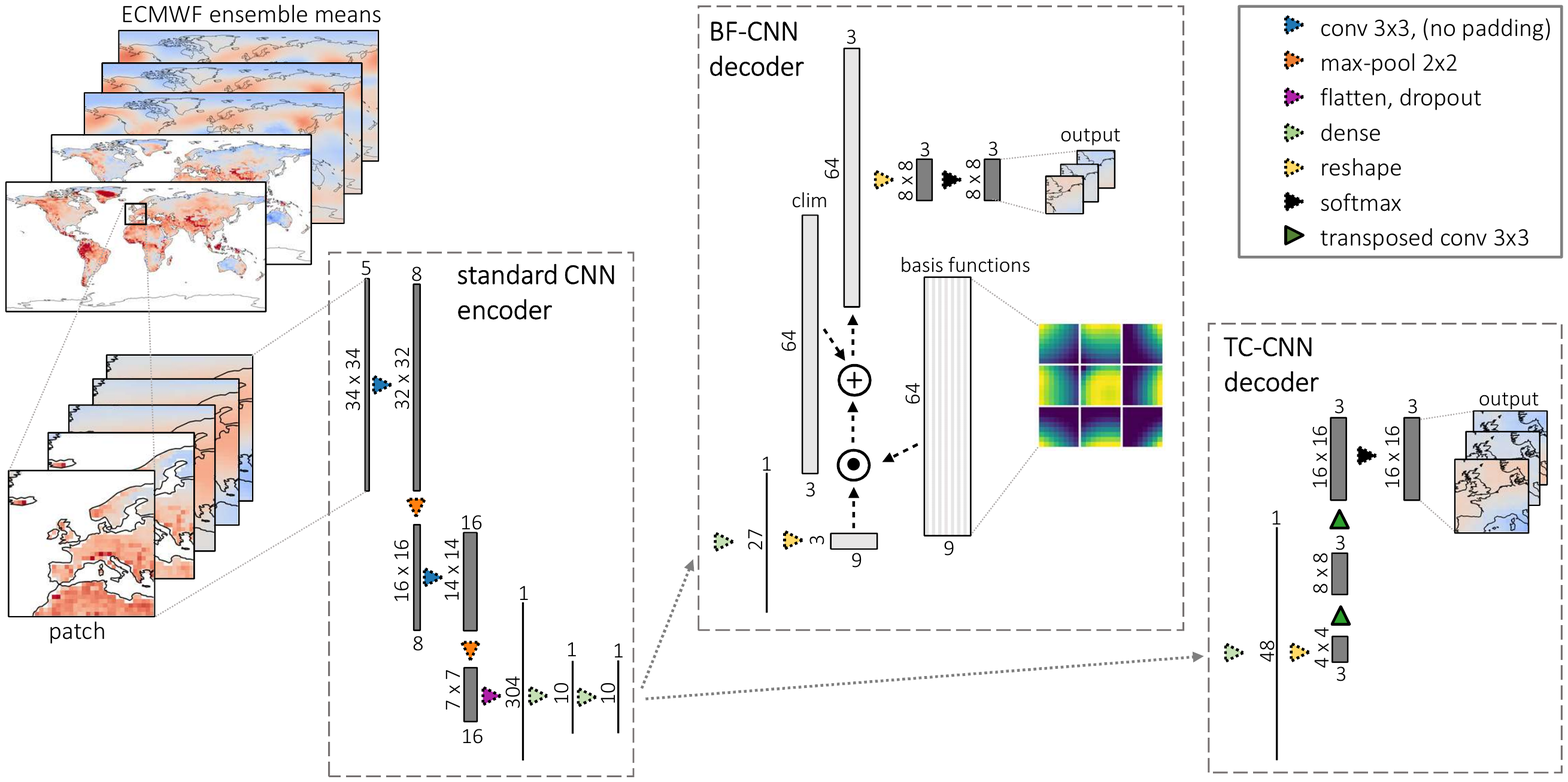}
		\caption{The architecture of the BF-CNN and the TC-CNN model for temperature (for precipitation, one dense layer less is used in the encoder and the input contains six channels).
			The two architectures share the same encoder, but use different strategies to derive a spatial prediction from the lower-dimensional representation. The decoder of the basis function CNN is shown in the middle on top, the corresponding decoder for the CNN with transposed convolutions on the right. The number of channels is denoted on top of each box, the size of the feature maps or the length of the vector is given on the left side of each box. Dark grey shaded boxes symbolize 3-dimensional matrices, while light grey boxes correspond to 2-dimensional fields, and the thin black lines denote 1-dimensional vectors.
		}
		\label{fig:schematicsCNNs}
	\end{figure}

	In addition to UNet-based models, we further explore ways of adapting a standard CNN architecture to enable spatial predictions. \citet{scheuererBasisfunc} propose a standard CNN that returns coefficient values for local, spatially smooth basis functions as output in order to create spatial forecasts for precipitation over California on sub-seasonal time scales. A direct application of this method to global data is infeasible since the coefficient matrix for an appropriate number of basis functions becomes prohibitively large for a global output domain at 1.5$^\circ$ resolution. 
	We therefore extend this approach to global data by splitting the original global map into multiple smaller domains, so-called patches, and train the basis-function CNN (BF-CNN) on these patches. 
	This technique is often used in image segmentation to augment training data and reduce the complexity of the model \citep[e.g.,][]{ciresan2012, imageSegmetationReview, patchwiseTraining}. Figure \ref{fig:schematicsCNNs} shows a schematic of this model architecture. 
	Note that the patches of input data will generally be larger than the corresponding output patches on which the models produce probability forecasts.
	
	We further propose a slightly more flexible variant of the BF-CNN approach by replacing the basis functions with transposed convolutional layers as depicted in Figure \ref{fig:schematicsCNNs}. These transposed convolutional layers directly yield predictions of category probabilities for all pixels within a considered output patch. Therefore, this model is more flexible but also has more trainable parameters. We abbreviate this model architecture with TC-CNN.
	
	The two models based on standard CNN architectures, namely BF-CNN and TC-CNN, are always trained patch-wise and produce separate predictions for each patch. The UNet architecture on the other hand can be trained either directly on global input data (global UNet) or patch-wise on smaller parts of the global forecasts (patch-wise UNet). We will abbreviate patch-wise with pw and append this abbreviation to the model names of the patch-wise trained models in figures and tables to improve clarity. In the following, we describe input and  model architecture of these different methods in more detail. 
	
	\subsection{Model inputs}
	Global ECMWF forecasts of multiple meteorological variables serve as input to our post-processing models. In addition to the the target variable forecasts, we use geopotential height forecasts at \SI{500}{hPa} and \SI{850}{hPa}, as well as sea-level pressure forecasts as inputs for post-processing temperature forecasts. For precipitation, total column water forecasts are considered in addition since we assume the forecast skill of total column water forecasts to be higher than for precipitation. The precipitation (re-)forecasts contain negative and missing values, in particular in drier regions. We set all negative precipitation accumulations to zero.
	
	Instead of using the whole ensemble as predictor for our post-processing models, we only use the ensemble mean. We do not use the ensemble spread or other measures of variability within the ensemble members since the benefit of using those as additional predictors has been found to be small in previous studies \citep[e.g.,][]{rasp_neural_2018,SchulzLerch2022}. 
	For providing tercile probability forecasts, the exact values of the target variable predictor forecasts are not relevant but relative values with respect to the boundaries between the three categories are more informative. Therefore, we compute the distance from the ensemble mean of the target variable forecast to the respective tercile category edges, which yields the distance to the upper and lower boundary of the middle tercile as two new predictors replacing the ensemble mean. 
	For the non-target variable predictors, we compute anomaly values for the ensemble mean with respect to a local and week-specific climatology (i.e., a separate climatology for each grid cell and week of the year). All predictors are further standardized with the grid cell-specific standard deviation. Before feeding the spatial fields to the ML models, we fill all missing values (i.e., all ocean grid cells and some very dry grid cells) with zeros. 
	
	\subsection{Architecture details for the UNet models}
	
	The UNet architecture is a CNN with an encoder-decoder architecture. The encoder part, responsible for decreasing the resolution of the input fields and for learning patterns, here consists of three blocks. The $i$\textsuperscript{th} block comprises two convolutional layers with $4 \cdot 2^i$ filters followed by a batch normalization and an average pooling layer. These encoder blocks are followed by a bottle-neck block that contains two convolutional layers with 64 filters and a batch normalization layer. The output of the bottle-neck is fed into the decoder part, that is responsible for the up-scaling of the low-resolution pattern to the original resolution, and also consists of three blocks. The $i$\textsuperscript{th} block (with $i$ from 3 to 1, i.e. in reversed order) starts with a transposed convolutional layer with $4 \cdot 2^i$  filters, whose output is then concatenated with the output of the batch-normalization layer of the $i$\textsuperscript{th} encoder block. The result of this so-called ``skip-connection'' is further processed by two convolutional layers with $4 \cdot 2^i$  filters and a batch normalization layer. One final convolutional layer with 3 filters with kernel size and stride equal to one and softmax activation function concludes the UNet architecture.
	Unless stated differently, all convolutional layers in the UNet architecture use a kernel size of 3 x 3, stride one, \textit{same} padding and exponential linear unit (ELU) activation. The transposed convolutional layers use the same parameters except that a stride of two is used.
	
	\subsection{Architecture details for the standard CNN models}
	We slightly adapt the model architecture of the basis-function CNN model proposed in \citet{scheuererBasisfunc} depending on the target variable. For precipitation we use the same encoder architecture as in \citet{scheuererBasisfunc}, for temperature we double the number of filters, resulting in the following encoder: 
	\begin{itemize}
		\itemsep0em 
		\item 3 x 3 convolutional layer with eight filters, stride two, no padding and ELU activation, 
		\item 2 x 2 max pooling layer
		\item 3 x 3 convolutional layer with 16 filters, stride two, no padding and ELU activation,
		\item 2 x 2 max pooling layer
	\end{itemize}
	In both cases, we use input patches of $34 \times 34$ grid cells.
	
	The coarse resolution two-dimensional fields obtained by these operations are flattened and dropout is applied with dropout rate of 0.4. One (for precipitation) or two (for temperature) dense layer with 10 nodes and ELU activation conclude the encoder part of the BF-CNN and the TC-CNN models. 
	
	The decoder part responsible for the up-scaling to the target domain differs between the BF-CNN and the TC-CNN models. For the BF-CNN another dense layer with ELU activation and 27 nodes is applied, with the number of nodes matching the number of basis functions (9) times the number of terciles (three). The resulting vector is reshaped into a $3 \times  9$ matrix and represents the coefficient values for the basis functions. 
	The nine entries per tercile are now mapped to the $8 \times 8$ output domain by multiplying with the nine basis functions ($64 \times 9$ matrix).  
	The basis functions are spatially smooth functions with circular bounded support (with a radius of 16 grid cells) that are uniformly distributed over the output domain, see \citet{scheuererBasisfunc} for details. 
	An illustration is included as part of Figure \ref{fig:schematicsCNNs}. 
	Following \citet{scheuererBasisfunc}, we add a vector of logarithmic climatological probabilities (equal to $1/3$ in our case) and reshape the output into a tensor of shape $8 \times 8 \times 3$. Applying the softmax function to this tensor across the third dimension yields a probabilistic forecast for terciles for a domain of $8 \times 8$ grid cells.
	
	The decoder part of the TC-CNN model is slightly simpler. The output  of the last dense layer of the encoder is passed through another dense layer with 48 ($3 \times 16$) nodes and ELU activation. This vector is reshaped into a $4 \times 4 \times 3$ tensor. Next, two $3 \times 3$ transposed convolutional layers are applied with three filters and stride two, ELU activation and \textit{same} padding, increasing the domain size from $4 \times 4$ over $8 \times 8$ to $16 \times 16$, matching the output domain of $16 \times 16$ grid cells. 
	Again, applying the softmax function across the third dimension of the output tensor yields a probabilistic forecast for terciles for a domain of $16 \times 16$ grid cells.
	The optimal size of the output domain (eight versus 16 grid cells) was found by hyperparameter optimization using cross-validation, as further described below. 
	
	\subsection{Model training}\label{technicalDetails}
	Our ML methods are implemented in Python using Keras \citep{chollet2015keras} and TensorFlow \citep{abadi2016tensorflow}. During model training, we minimize the categorical cross-entropy with the Adam optimizer \citep{adam2015}. The training and hyperparameter optimization is done with 10-fold cross-validation on 20 years of training data (2000--2019). The 10 folds consist of two consecutive years each, i.e., 2000 and 2001, 2002 and 2003, and so on. Each fold is used once as validation fold, while the remaining nine folds are used for model training.  The pre-processing is done separately for training and validation data of each split to prevent data leakage. The final prediction for the test year 2020 is obtained by averaging over the predictions of the 10 models obtained by 10-fold cross-validation. 
	
\begin{figure}
		\centering
		\includegraphics[width=0.6\textwidth, trim={0 0 0 0},clip]{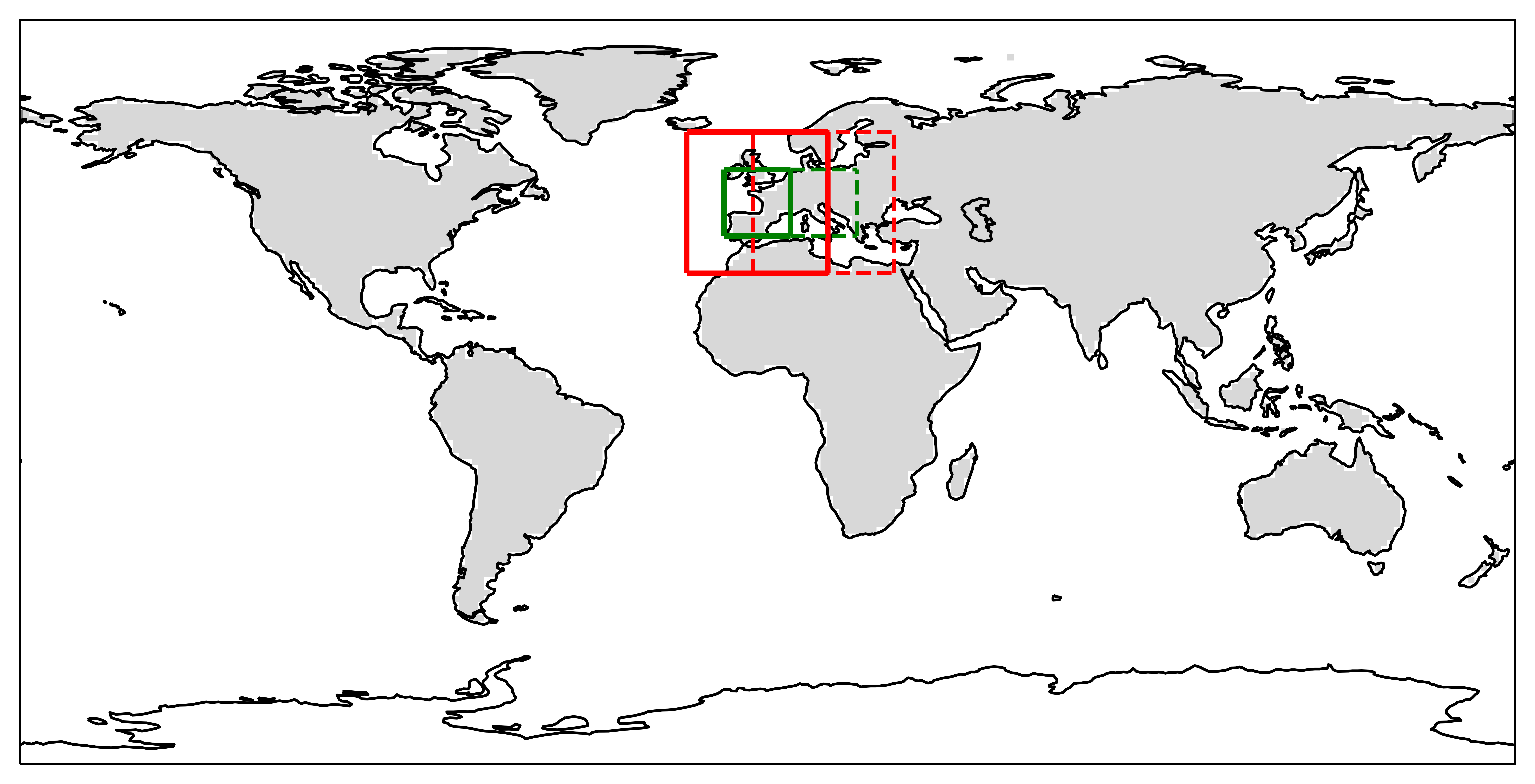}
		\caption{Visualization of example input patches (red) and the respective output patches (green) for the patch-wise models. The solid and dashed squares represent two different patches used for creating a connected spatial prediction for a larger domain. }
		\label{fig:predictPipeline}
	\end{figure}

	For the patch-wise models, the data preparation also includes generating patches, i.e., selecting small quadratic domains from the global input fields of, for example, eight grid cells in longitude and latitude direction each. As indicated in Figure \ref{fig:predictPipeline}, input patches (in red) are chosen slightly larger than output patches (in green) to improve the prediction at the patch edges. We allow for overlapping training patches to increase the number of training samples and use a patch stride of 12, which means that once we have selected a patch, we move 12 grid cells east or south before selecting the next patch. The amount of overlap is different for every architecture depending on the size of the output patch. Not all patches sampled in this way are also useful for training, since no ground truth data is available over the ocean. 
	Therefore, we discard all patches with more that 50\% missing values in the output patch.
	For the prediction, the output patches are defined without overlap (Figure \ref{fig:predictPipeline}, dashed and solid lines) and the separate predictions for the individual patches are stitched together to obtain a global prediction. 
	As discussed earlier, the predictor patches are larger than the output patches. 
	To make the output patches cover the whole globe, we therefore have to pad the global input fields before patch selection. 
	The padding together with the specific patch sizes for the different model architectures is detailed in Appendix \ref{appendix:padding}. 
	Note that we also pad the input of the globally trained UNet to avoid large discontinuities at the date line and the poles. 
	
	Hyperparameters related to model training vary between global and patch-wise training and the different model architectures. The global UNet is trained for 50 epochs with a batch size of 16. The patch-wise models are trained for 20 epochs with a larger batch size of 32 or 64 since more training samples are available. We have roughly 1\,000 global forecasts, resulting in more than 30\,000 valid samples for the patch-wise models. 
	We always use early stopping but adapt the patience according to the total number of training epochs, resulting in a patience of 10 and 3 for the global and the patch-wise models, respectively. 
	Since the precipitation forecasts are much less accurate and the corresponding ground truth data is much more noisy than for temperature, the post-processing models have more difficulties to learn from precipitation data. 
	We therefore introduce delayed early stopping and adjust the learning rate for the precipitation models. The training configuration is detailed in Appendix \ref{appendix:earlystop}.
	
	The loss function is averaged over all grid cells of the entire globe (for the global UNet) or over the output patch (for the patch-wise models). While it is standard practice in meteorological applications to weigh the contribution of each grid cell with the corresponding area fraction of the grid cell (cosine of the latitude), ML architectures operating on image-like data usually do not use any pixel (grid cell) weighting scheme when computing the loss function on the entire image. However, ML models for data-driven weather prediction often adapt model architecture and data processing to account for the spherical nature of atmospheric data \citep[e.g.,][]{weyn_improving_2020}.
	We therefore here adapt the standard categorical-cross entropy loss to account for the grid-cell weight during training. This modification can be used for both, patch-wise and global training. For the patch-wise training it leads to differently weighted patches, depending on the patch location. The effect of introducing the grid cell weights on the predictive performance is discussed in detail in Section \ref{resWeighted}.

	\section{Results}\label{sec:results}
	
	\subsection{Forecast evaluation setup}
	Following the setup of the S2S AI Challenge, we evaluate the proposed post-processing methods based on the ranked probability skill score (RPSS) for the test set (year 2020), using the climatological forecast as a reference. The RPSS is based on a strictly proper scoring rule and introduced in detail in Appendix \ref{evalMethods}. It is positively oriented, i.e., larger values indicate better forecasts. Following the challenge configuration, only land grid cells are considered for the computation of the spatially aggregated RPSS scores since no observations are available for sea grid cells. 
	Antarctica is not included in the aggregated RPSS. For precipitation, also very dry regions (lower tercile edge smaller than \SI{0.01}{mm/m^2} are not taken into account. 
	For each grid cell, the average ranked probability score (RPS; see Appendix \ref{evalMethods} for details) of the model predictions and the climatological reference forecast (described in Section \ref{benchmarks}) is computed by aggregating over the 53 Thursday forecasts in 2020, which form the test data set of the S2S AI Challenge. From these grid cell-wise RPS values, we compute a grid cell-wise RPSS, which is then averaged over the target domain (weighted by the area fraction of each grid cell). 
	
	\begin{figure}
		\centering
		\includegraphics[width=1\textwidth,trim={0 0 0 1cm},clip]{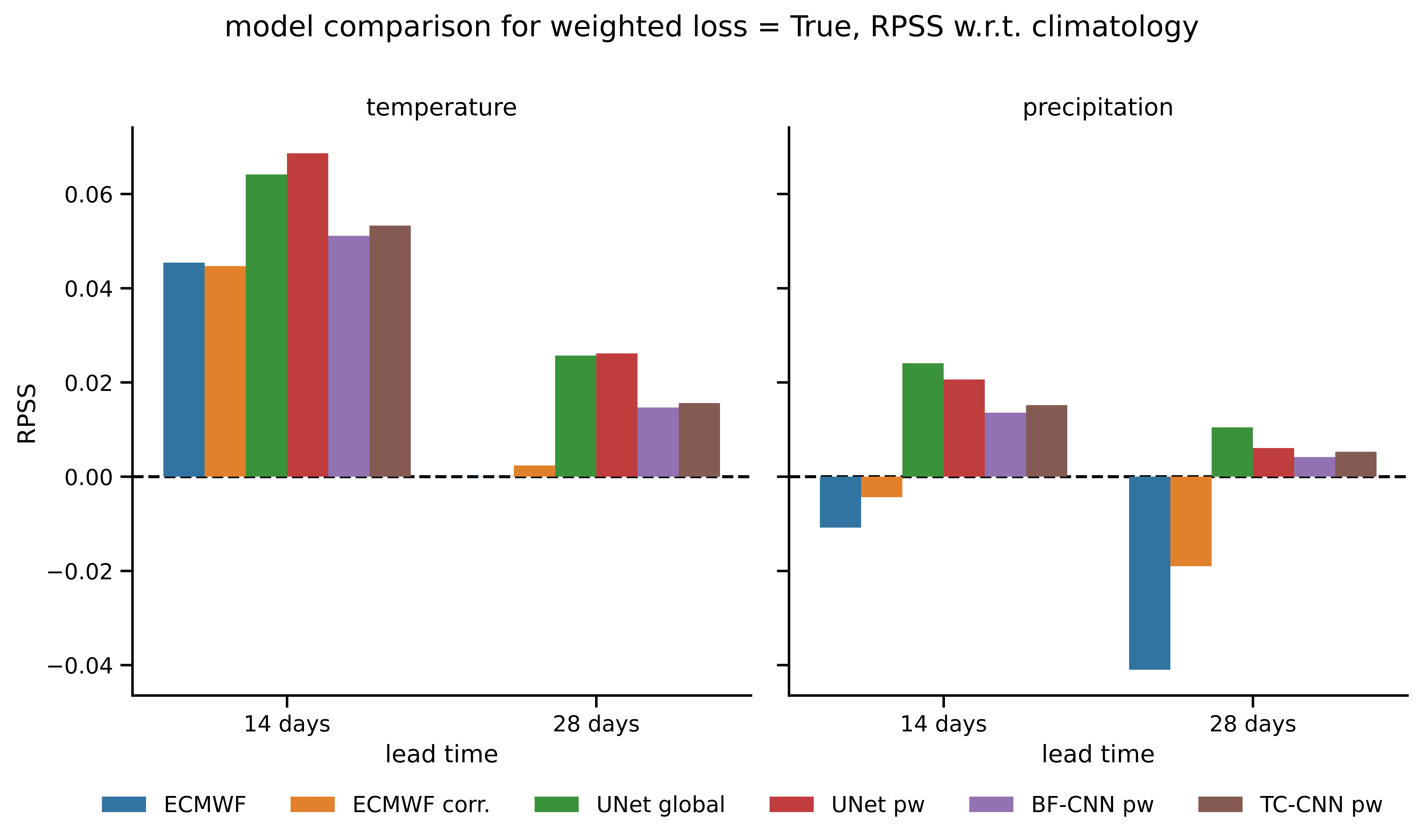}
		\caption{Globally aggregated RPSS for the test year 2020. 
			RPSS values larger than zero indicate that forecasts are better than climatology (i.e., random guessing). The RPSS of the ECMWF temperature forecasts for a lead time of 28 days is positive, but orders of magnitudes too small to be visible in the figure.
		}
		\label{fig:summaryBarPlot}
	\end{figure}
	
	\subsection{General results}
	Figure \ref{fig:summaryBarPlot} summarizes the globally averaged RPSS of the ML methods and the two ECMWF benchmarks, consisting of the benchmark provided by the challenge organizers and our corrected ECMWF baseline. All proposed post-processing methods show positive skill when compared to climatological forecasts, and generally outperform the ECMWF baseline forecasts as indicated by the higher RPSS. The largest improvements over the ECMWF baselines are achieved for the precipitation forecasts, where all post-processing methods are able to turn the highly non-skilful forecasts into skilful forecasts ($\text{RPSS} > 0$). For temperature, the relative improvement is smaller, in particular for the shorter lead time of 14 days (i.e., weeks 3-4). 
	Further, the positive effects of the correction of the ECMWF baseline forecasts discussed in Section \ref{benchmarks} are clearly indicated by the substantial improvement in terms of RPSS.
	
	The UNet architectures achieve the best scores for all four test cases (temperature and precipitation for weeks 3-4 and weeks 5-6). We attribute this to the fact that the UNet architecture is very well suited for the task of splitting the global forecasts into different regions (above, near and below normal), since its architecture was specifically developed for image segmentation. This allows us to train a well performing model with as little as roughly 1\,000 global forecasts. Overall, the global UNet yields the best results with a global RPSS of 3.1\% averaged over all four test cases. For temperature for weeks 3-4, the global UNet is outperformed by the patch-wise UNet. 
	The BF-CNN and TC-CNN models show fairly similar predictive performances, with slight advantages for the more complex, but also more flexible TC-CNN approach.
	
	All figures and numbers show the RPSS of the average prediction obtained from 10 model runs based on the cross-validation procedure applied for training. The skill of this mean prediction is always higher than the expected skill of the individual model runs. Model averaging yields the largest benefit for the global UNet with a relative improvement of almost 30\% compared to the average skill of the individual model runs. The skill of the other approaches improves by 10 to 17\%. 
	
	\begin{figure}[p]
		\centering
		\includegraphics[width=\textwidth,trim={0 4.05cm 0 0cm},clip]{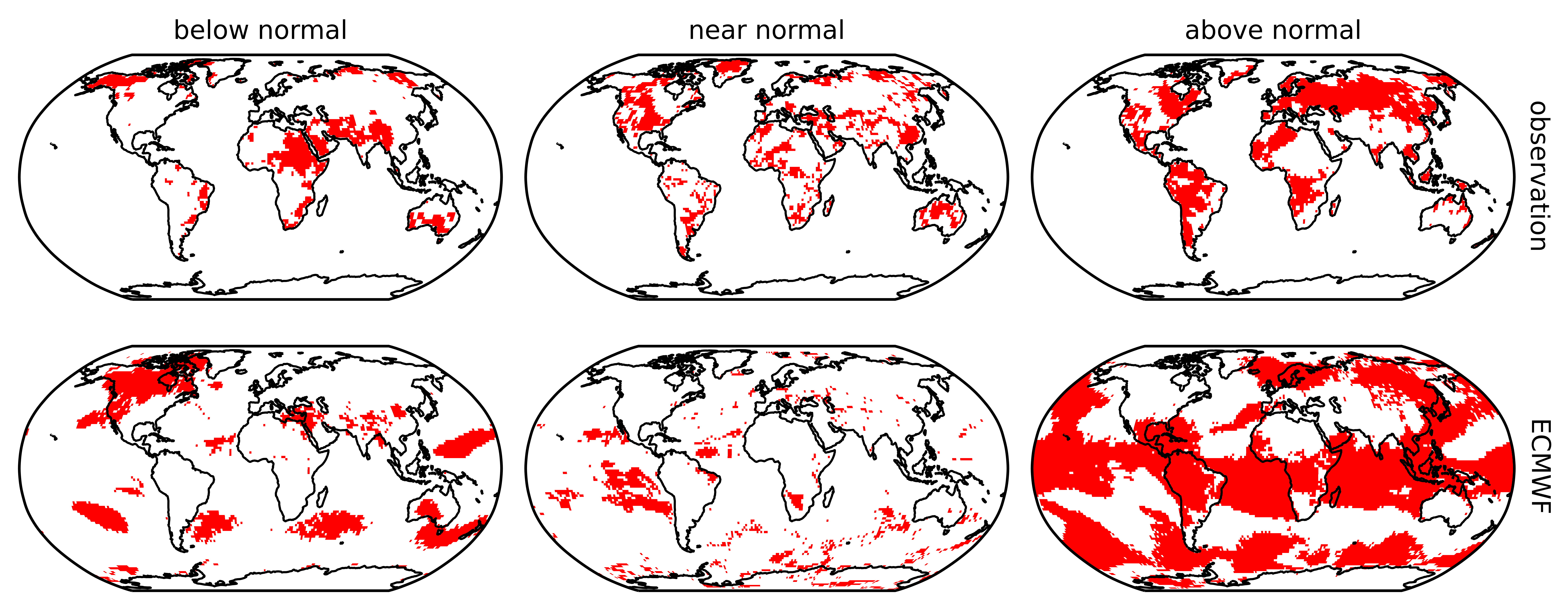}
		\includegraphics[width=0.13\textwidth,trim={11.5cm 7.85cm 0.55cm 1.5cm},clip]{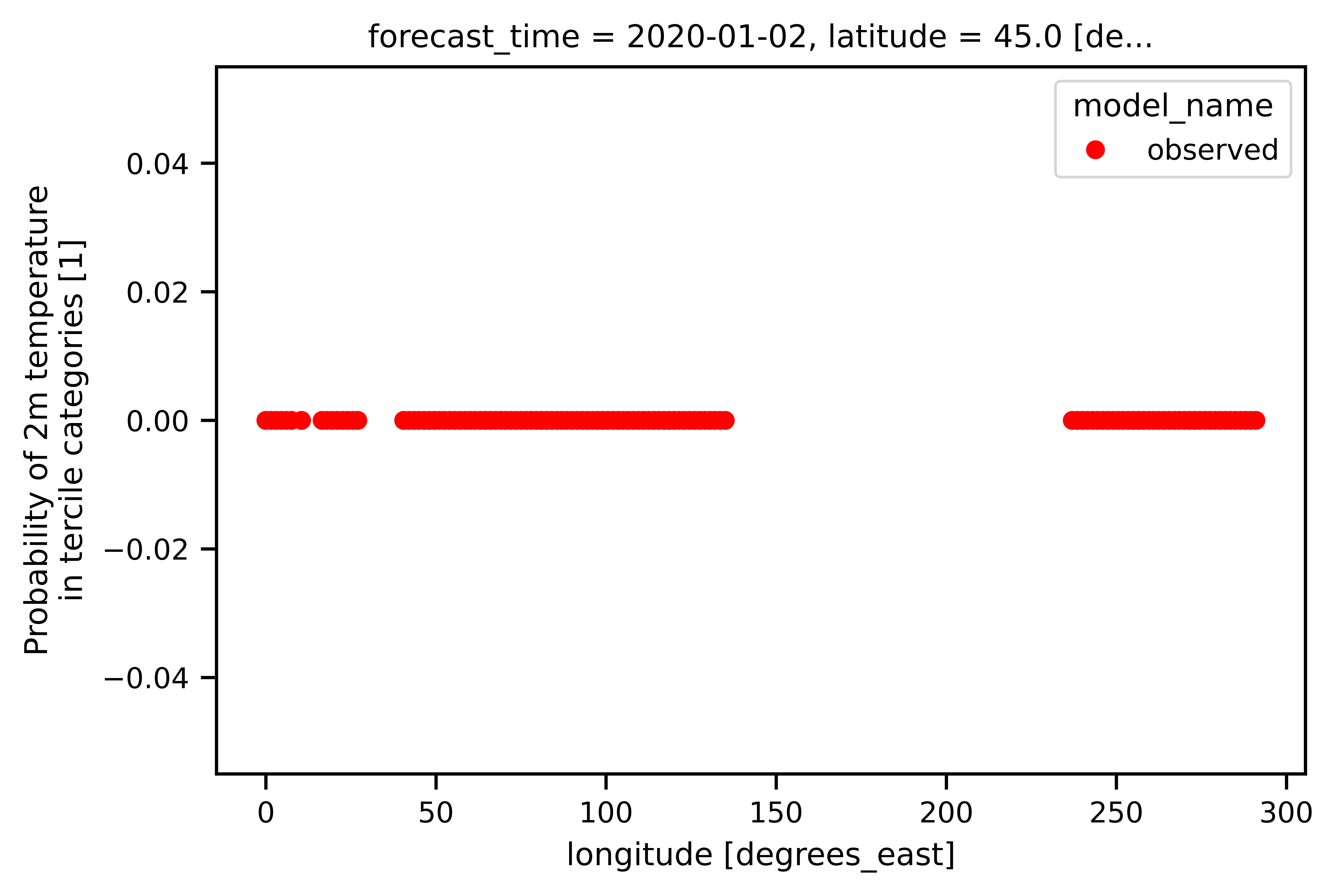}\hspace*{8cm}
		
		\vspace{0.2cm}
		\hfill
		\includegraphics[width=\textwidth,trim={0 0 0 0.1cm},
		clip]{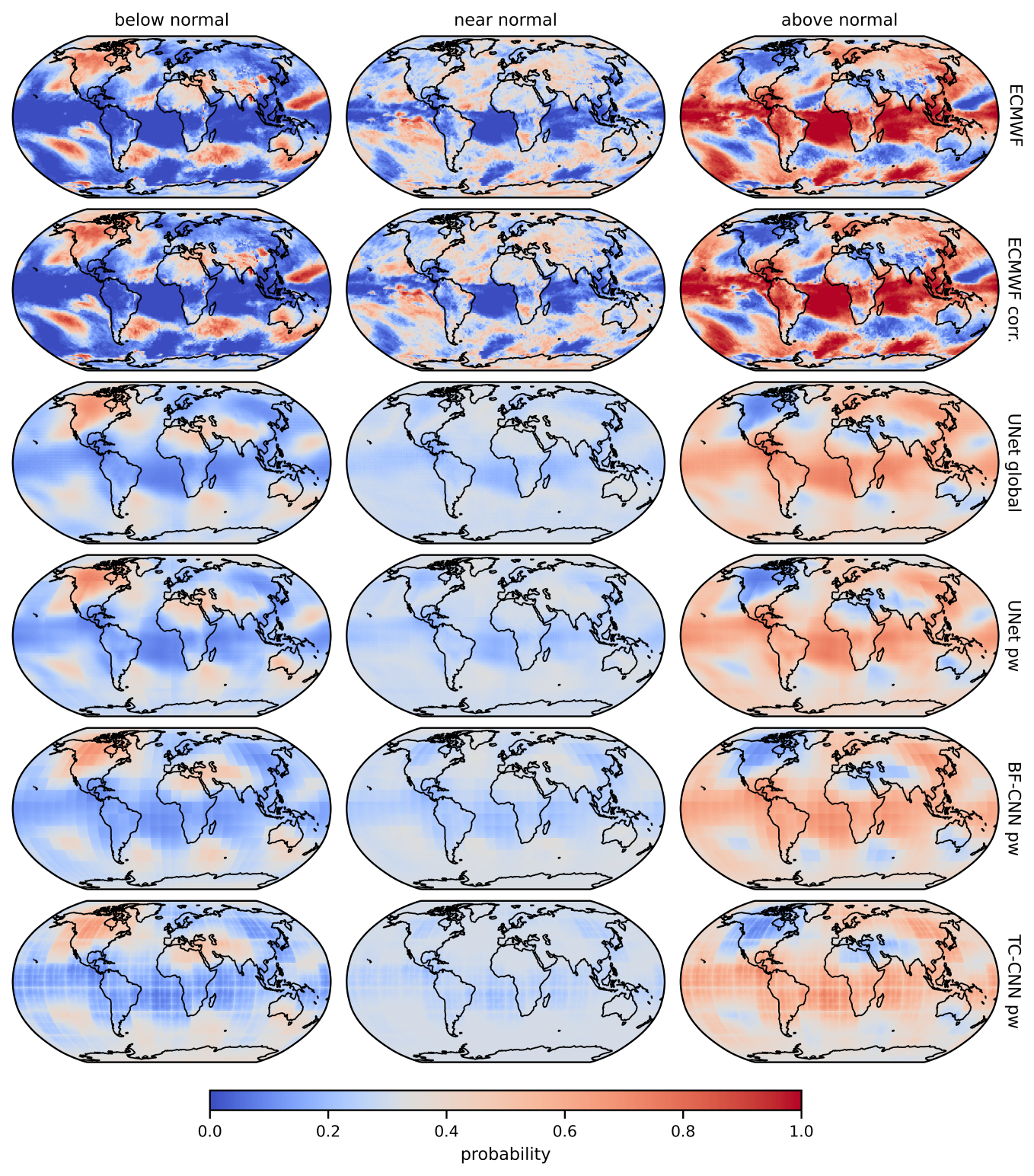}
		\caption{Example predictions for temperature with a lead time of 14 days issued on 2 January 2020. In the top row, the corresponding observations are displayed, followed by the predictions of the two benchmark methods and the four post-processed forecasts.
		}
		\label{fig:exampleForecasts}
	\end{figure}

	Figure \ref{fig:exampleForecasts} shows the post-processed predictions for an example forecast issued in January 2020 for temperature for weeks 3-4. Additional example forecasts for precipitation are available in Appendix \ref{appendix:plots}, and forecasts for weeks 5-6 are provided in the supplemental material\footnote{available at \url{https://github.com/HoratN/pp-s2s/blob/main/Supplementary_Material.pdf}}. While all methods generally agree on the spatial structure of the temperature distribution, there are visible differences between the methods due to the different training modes and architectures. The patch-wise models capture the within-patch gradient properly for most patches, leading to a reasonably smooth forecast. However, the patch edges are still visible as small discontinuities. This could potentially be reduced by choosing a larger margin between input domain and output domain. It is interesting to note that the forecasts over the ocean also look reasonable even though no ground truth data was available there. 
	
	The post-processed forecasts are generally much smoother and less sharp than the ECMWF benchmarks. Further, the forecasts for weeks 5-6 are less sharp than the week 3-4 forecasts. Since ensemble forecasts from numerical weather prediction models have been demonstrated to often be overconfident  \citep[e.g.,][]{vannitsem_statistical_2021}, it is not surprising that post-processing reduces the forecast sharpness in order to improve calibration, as discussed in more detail in Section \ref{calibration}. 
	The reduced sharpness of the post-processed forecasts also is in line with the findings of \citet{Vigaud2017MMEprecip} who post-processed sub-seasonal precipitation forecasts. 
	The probabilities for the near normal category are very close to the climatological probability of 1/3, in particular for the post-processed forecasts. The scattered/intermittent nature of the near normal tercile and the small differences between the boundaries of the upper and lower terciles likely favoured conservative forecasts for the middle tercile. This issue is discussed in detail in Section \ref{criticalTerciles}.

	\subsection{Spatial verification}
\begin{figure}[p]
		\centering
		\includegraphics[width=0.75\textwidth,trim={0 0 0 0cm},clip]{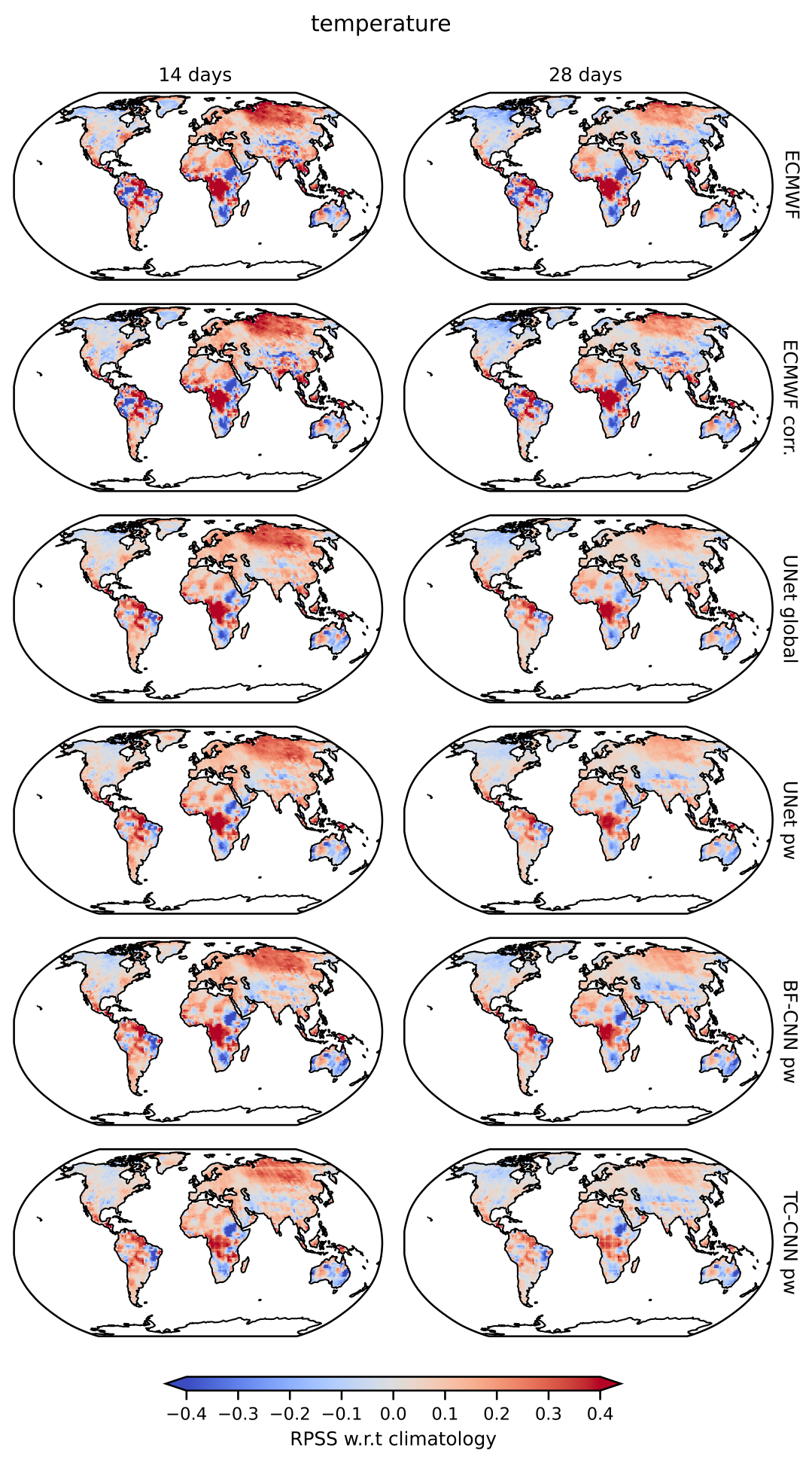}
		\caption{Grid cell-wise RPSS for temperature for 2020.
			RPSS values larger than zero indicate that forecasts are better than climatology.}
		\label{fig:spatialRpssT2m}
	\end{figure}

\begin{figure}[p]
		\centering
		\includegraphics[width=0.75\textwidth,trim={0 0 0 0cm},clip]{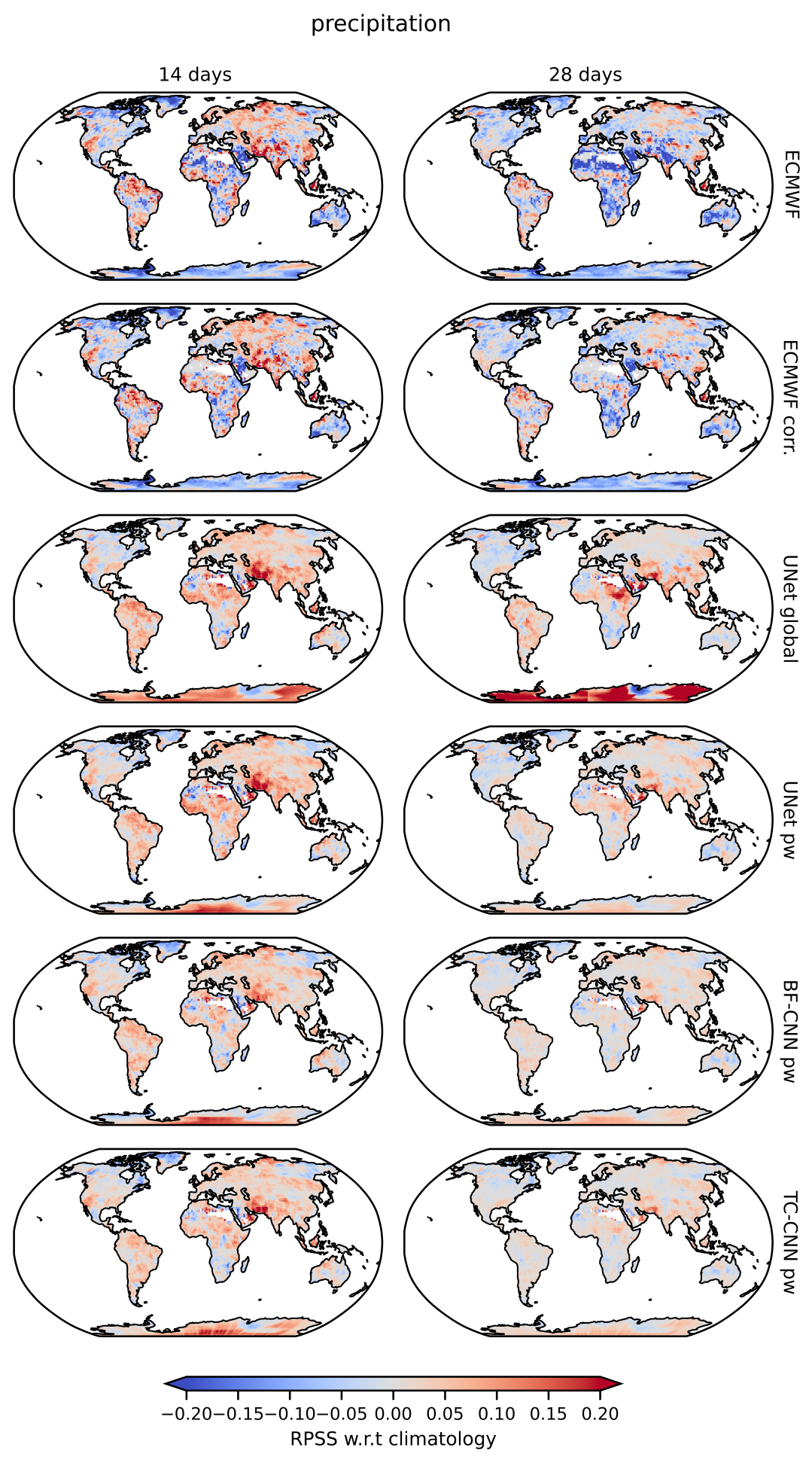}
		\caption{Grid cell-wise RPSS for precipitation for 2020.
			RPSS values larger than zero indicate that forecasts are better than climatology. Very dry grid-cells are omitted from the evaluation (e.g.,\ parts of the Saharan Desert).}
		\label{fig:spatialRpssTp}
	\end{figure}

	For evaluating the spatial differences in model performance, we on the one hand define three regions, the northern extra-tropics, abbreviated as NH, that correspond to 90$^{\circ}$-30$^{\circ}$N, the tropics corresponding to 30$^{\circ}$N-30$^{\circ}$S, and the southern extra-tropics, abbreviated as SH and covering 30$^{\circ}$-60$^{\circ}$S. On the other hand, we also present global grid cell-wise RPSS maps in Figures \ref{fig:spatialRpssT2m} and \ref{fig:spatialRpssTp} to show the fine spatial details.
	
	\begin{table}
		\caption{RPSS averaged over the northern extra-tropics (NH; 90$^{\circ}$-30$^{\circ}$N), the tropics (30$^{\circ}$N-30$^{\circ}$S), and the southern extra-tropics (SH; 30$^{\circ}$-60$^{\circ}$S). The global domain consists of all three regions mentioned before, and corresponds to the values shown in Figure \ref{fig:summaryBarPlot}. The grey shading highlights the best performing model for each task. }
		\begin{adjustwidth}{-1in}{-1in}
			\centering
			\begin{tabular}{ll|rrrr|rrrr}
				\toprule
				&  & \multicolumn{4}{c}{temperature} & \multicolumn{4}{c}{precipitation} \\
				&  &  global &     NH & tropics &     SH& global &     NH & tropics &     SH \\
				lead time & model &        &        &         &        &        &        &         &        \\
				\midrule
				14 days & ECMWF &  0.045 &  0.052 &   0.074 &  \cellcolor{black!15}-0.001 & -0.011 & -0.012 &   0.003 & -0.004 \\
				& ECMWF corr. &  0.045 &  0.053 &   0.068 & -0.013 & -0.004 & -0.012 &   0.017 & -0.003 \\
				& UNet global &  0.064 &  0.074 &   0.088 & -0.012 &   \cellcolor{black!15}0.024 &   \cellcolor{black!15}0.014 &    \cellcolor{black!15}0.051 &   \cellcolor{black!15}{0.024} \\
				& UNet pw &  \cellcolor{black!15}{0.069} &  \cellcolor{black!15}{0.078} &   \cellcolor{black!15}{0.089} & -0.011 &   0.021 &  0.009 &   0.048 &  0.022 \\
				& BF-CNN pw &  0.051 &  0.066 &   0.064 & -0.031 &  0.014 &  0.005 &   0.031 &  0.022 \\
				& TC-CNN pw &  0.053 &  0.067 &   0.062 & -0.008 &  0.015 &  0.005 &   0.037 &  0.016 \\
				\midrule
				28 days & ECMWF &  0.000 & -0.002 &   0.045 & -0.052 & -0.041 & -0.032 &  -0.054 & -0.015 \\
				& ECMWF corr. &  0.002 &  0.001 &   0.046 & -0.047 & -0.019 & -0.023 &  -0.005 & -0.014 \\
				& UNet global &   \cellcolor{black!15}{0.026} &  0.021 &    \cellcolor{black!15}{0.066} & -0.031 &   \cellcolor{black!15}{0.010} &   \cellcolor{black!15}{0.002} &    \cellcolor{black!15}{0.035} &   {0.010} \\
				& UNet pw &   \cellcolor{black!15}{0.026} &   \cellcolor{black!15}{0.024} &   0.058 & -0.042 &  0.006 & -0.001 &   0.024 &  0.006 \\
				& BF-CNN pw &  0.015 &  0.015 &   0.044 & -0.059 &  0.004 &  0.001 &   0.013 &   \cellcolor{black!15}{0.011} \\
				& TC-CNN pw &  0.016 &  0.016 &   0.040 &  \cellcolor{black!15}{-0.024} &  0.005 &  0.001 &   0.017 &  0.007 \\
				\bottomrule
				
			\end{tabular}
		\end{adjustwidth}
		\label{tab:regional_rpss}
	\end{table}
	
	For temperature, the ECMWF baselines already achieve good skill over the tropics (for both lead times) and the northern hemisphere (for weeks 3-4) as shown in Table \ref{tab:regional_rpss}. Our UNet-based post-processing methods further improve these forecasts. The largest skill is reached over Russia and parts of equatorial Africa and America, as well as over south-east Asia (see Figure \ref{fig:spatialRpssT2m}). The largest negative skill can be observed in the tropics, but also southern Africa and Australia contain regions which are poorly forecasted. 
	The skill of the two standard CNN architectures, BF-CNN and TC-CNN, is partly deteriorated by patch-edge artifacts.
	For precipitation, the picture is less clear since the regions with positive skill are more scattered than for temperature (see Figure \ref{fig:spatialRpssTp}) even though post-processing yields smoother skill maps. Striking is the high skill over Antarctica, where the post-processing models presumably were able to correct the bias in the ECMWF ensemble mean. 
	
	Certainly, one year of weekly forecasts is not enough to make conclusive statements in what regions the ECMWF forecasts and the post-processed predictions are skillful. Nevertheless, the overall improvement by post-processing is clearly demonstrated by the increase of the area with positive forecast skill compared to the benchmark methods.

	\subsection{Forecast calibration}\label{calibration}
	
	The overall aim of probabilistic forecasting is to maximize the sharpness of the predictive distribution, subject to calibration \citep{GneitingEtAl2007}. The example predictions in Figure \ref{fig:exampleForecasts} clearly indicate that the post-processed forecasts are less sharp than the ECMWF baseline, calling for an assessment of calibration. While the calibration of binary (dichotomous) categorical forecasts can readily be checked with reliability diagrams \citep{murphy_reliability_1977, wilks2011}, evaluating the calibration of a probabilistic three-category forecast is less straightforward. \citet{wilks_calibration_2013} proposes a so-called calibration simplex, an extension of the well-known reliability diagram to the tree-category case. Based on the R package \textit{CalSim} \citep{resin2021calsim}, we created calibration simplices for the corrected ECMWF baseline and all post-processing methods. 
	
\begin{figure}
		\centering
		\hspace{-1.5cm}\begin{tabular*}{\linewidth}{ccc}
			\footnotesize{\textsf{ECMWF corr.}} & \footnotesize{\textsf{UNet global}} &\vspace{0.1cm}\\
			
			\includegraphics[width=0.49\textwidth,trim={1.3cm 1.9cm 3.1cm 2.3cm},clip]{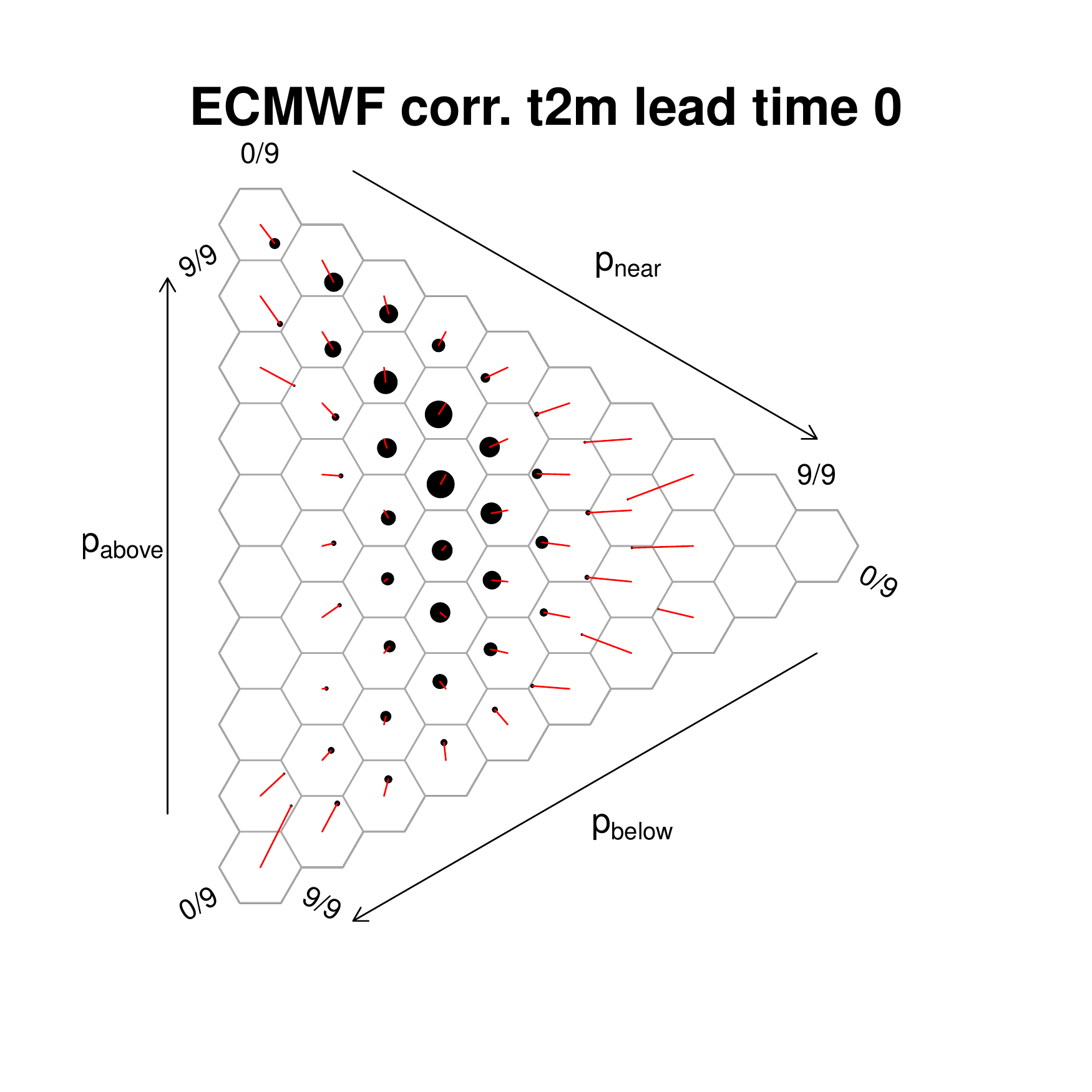}&
			\subfloat{
				\includegraphics[width=0.49\textwidth,trim={1.3cm 1.9cm 3.1cm 2.3cm},clip]{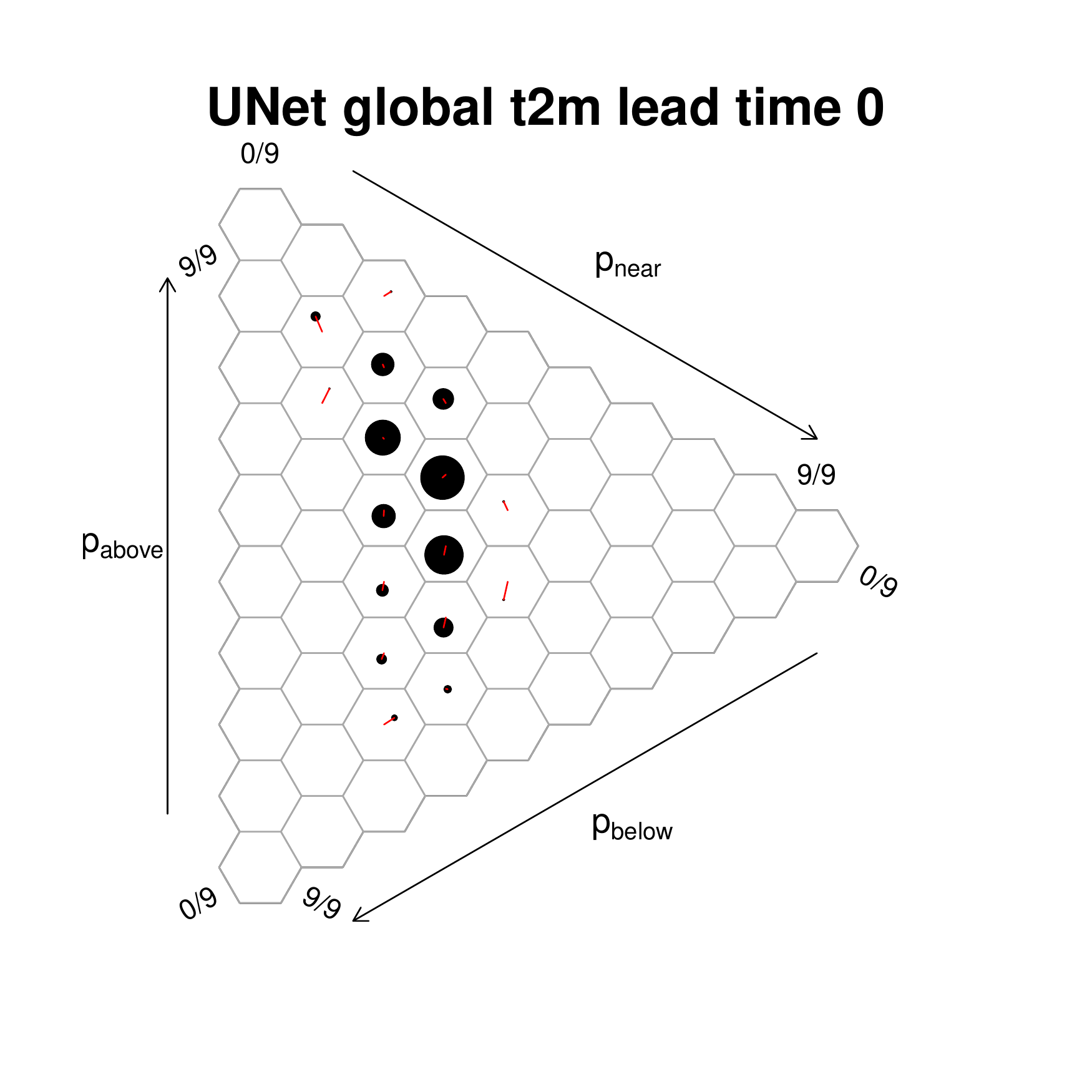}}&
			\subfloat{
				\raisebox{1.6in}{\rotatebox[origin=t]{-90}{\footnotesize{\textsf{temperature}}}}}
			\\
			
			\includegraphics[width=0.49\textwidth,trim={1.3cm 1.9cm 3.1cm 2.3cm},clip]{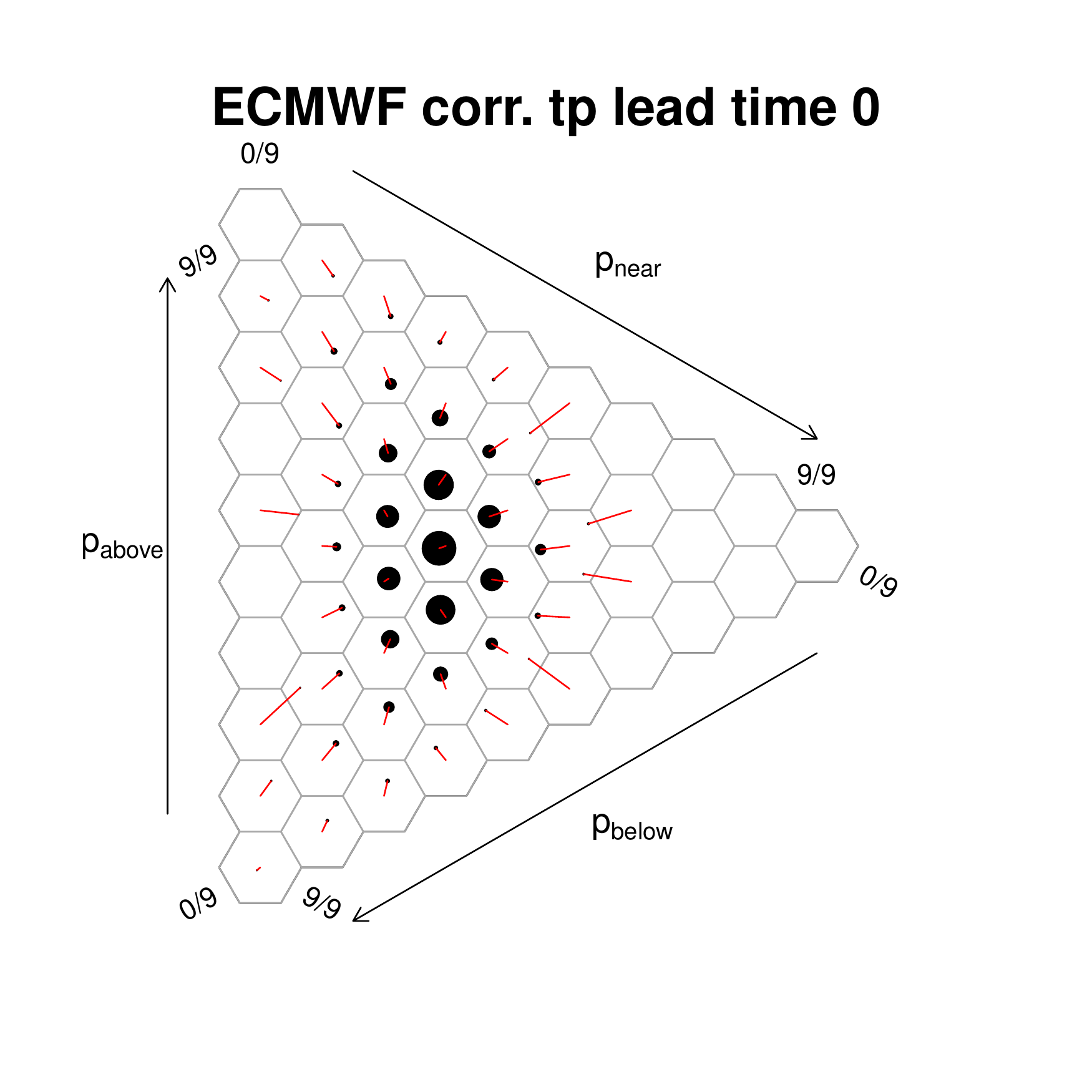}&
			\includegraphics[width=0.49\textwidth,trim={1.3cm 1.9cm 3.1cm 2.3cm},clip]{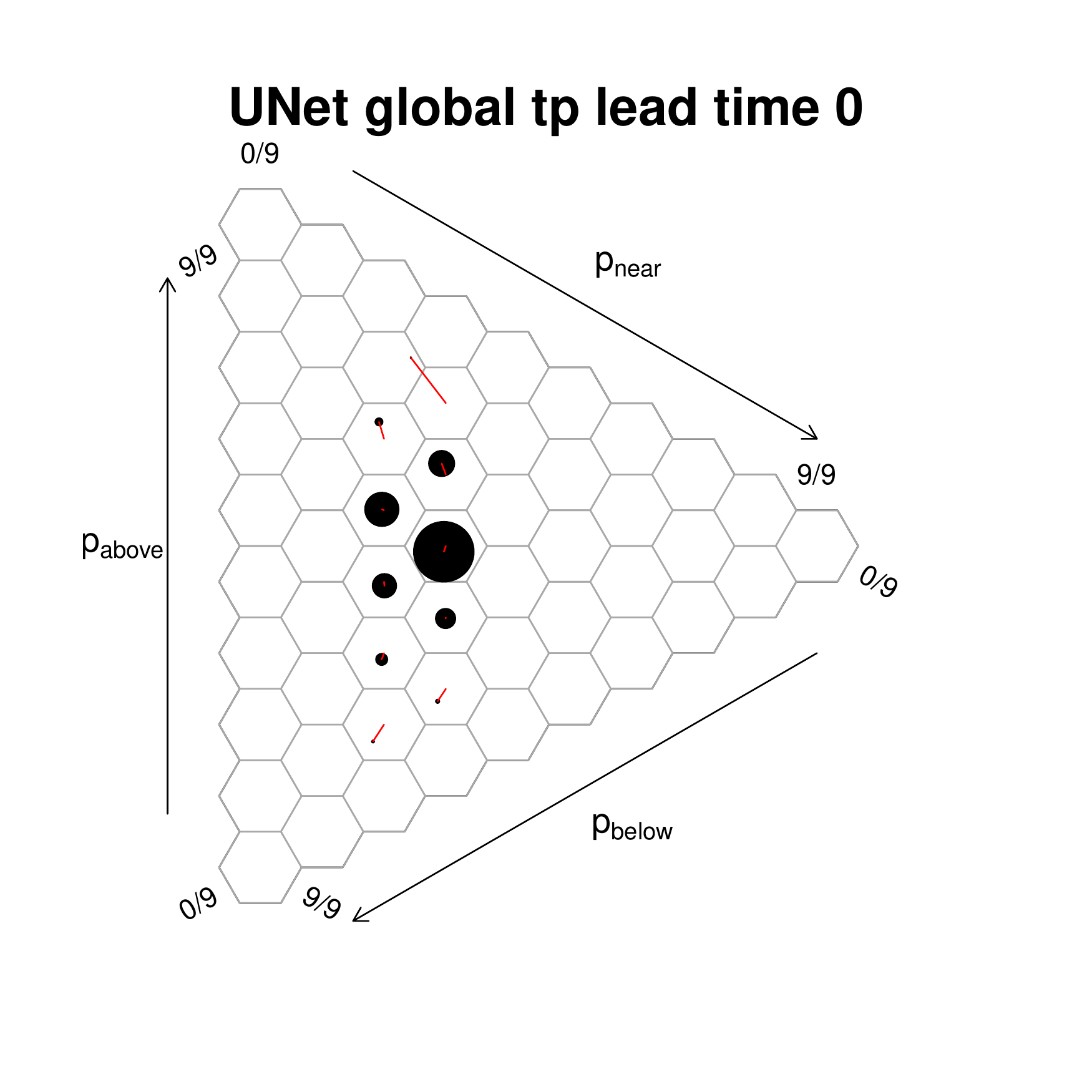}&
			\subfloat{
				\raisebox{1.6in}{\rotatebox[origin=t]{-90}{\footnotesize{\textsf{precipitation}}}}}\\
		\end{tabular*} 
		
		\caption{Calibration simplices for temperature and precipitation of the corrected ECMWF baseline and the global UNet. The simplices are based on all forecasts for 2020, but only consider land grid-cells north of 60$^{\circ}$S, i.e., omit Antarctica.}
		\label{fig:calibration_simplex}
	\end{figure}

	Since the simplices for the different post-processing approaches look very similar we only show results for the best performing model in terms of RPSS, the global UNet, in Figure \ref{fig:calibration_simplex}. As the reliability diagram, the calibration simplex depicts the predicted probabilities and the conditional frequency of the observed category. For each of the three terciles, forecast probabilities are binned in 10 bins (from 0/9 to 9/9) and the size of the dots in the hexagons represents the fraction of probability vectors falling in that ``three-dimensional'' bin. The displacement of the dot with respect to the center of each hexagon (indicated by the red lines) shows the miscalibration error. 
	The closer the dot is located to the center of the respective hexagons, the better calibrated the forecast.
	
	For both target variables the corrected ECMWF baseline produces too sharp forecasts, indicated by the substantial displacement of the dots towards the center of the calibration simplex. By contrast, the post-processing methods yields well-calibrated forecasts with appropriate sharpness given the limited predictability. Since the input to the post-processing models only consists of ensemble mean quantities, this indicates that the ML models are well able to correctly quantify forecast uncertainty from deterministic predictor information alone. 
	This finding is in line with results from the post-processing literature \citep[e.g.,][]{rasp_neural_2018}, as well as \citet{sacco_evaluation_2022} who investigate how well NNs can estimate forecast uncertainty related to model error and initial conditions and conclude that neural networks can reliably estimate forecast uncertainty. 
	
	All temperature forecast correctly predict the above normal category more often than climatologically expected. This either indicates that 2020 was exceptionally warm or is simply a signal of climate change, as was also observed by \citet{wilks_calibration_2013} in his case study. 
	The probabilities for the near normal category are particularly close to the climatological value of 1/3, which can be seen from the vertical alignment of the dots belonging to the 4/9 bin of the near normal tercile.
	
	\subsection{Comparison to S2S AI Challenge submissions}
	
	Overall, the post-processing methods presented here would (retrospectively) place in the top three of the S2S AI Challenge submissions (Table \ref{tab:regional_rpss} and \cite{s2saiChallenge}).\footnote{
		Note that we had submitted forecasts from a model that was a much simpler variant of the BF-CNN approach described here to the S2S AI Challenge which placed fourth.
	} For most target variable and forecast horizon combinations, our models show a solid performance and would place second or third. That said, the distance to the winning team is partially fairly large. This is not surprising since the winning team in the challenge proposed a complex mixture model based on locally corrected forecasts from ECMWF and other forecasting centers, a CNN-based correction of the ECMWF forecasts, and climatology. The multi-model approach presumably leads to large skill improvements, independent of the applied post-processing methods \citep{doblas-reyes_rationale_2005, Vigaud2017MMEprecip, li2017MME}.  
	
	A key distinguishing feature of our models, in particular of the global UNet approach, is that they do not rely on local or regional information only, but rather utilize global input information to produce global probabilistic predictions. 
	On the one hand, this leads to a lower performance for some regions with for example very specific temperature biases and potentially lower number of grid cells, such as the southern extra-tropics or the tropics. On the other hand, the use of spatial information might be the reason why our models perform very well for precipitation and achieve even similar performance to the winning team's predictions. In the tropics, our UNet models outperform all challenge predictions for precipitation.

	\subsection{Effects of using a weighted loss function for model training}\label{resWeighted}
	
	Overall, the absolute differences between the models trained without and with weighted loss are small. This implies that whether grid cell values are weighted during training according to their grid-cell size or not, does not yield entirely different post-processing models. 
	We compare the forecast skill of the models trained with weighted loss to the ones trained without by computing the RPSS of the weighted loss models using the unweighted models as reference,
	\begin{equation*}
	\text{RPSS}_\text{weighted} = \frac{\text{RPS}_\text{unweighted} - \text{RPS}_\text{weighted}}{\text{RPS}_\text{unweighted} - \text{RPS}_\text{opt}},
	\end{equation*}
	where $\text{RPS}_\text{opt} = 0$.
	
	The corresponding RPSS values are shown in Table \ref{tab:weighted}. It can be seen that the weighting improves the forecast skill the most in the tropics, since errors in this area influence the overall loss more strongly with the weighting. When using the weighted loss for the patch-wise models, patches from higher latitudes influence model training less than patches from equatorial regions. The weighting improves the predictions of the global UNet model the most, also having a positive effect on extra-tropical scores for temperature. For the patch-wise models the signal is less clear and changes in skill are mostly small. Hence, the weighted loss seems to influence the patch-wise training less, likely because the individual patches are too small for the weighting to have any real effect and the weighting of the patches as a whole happens only within training batches. 
	We conclude, that for the patch-wise models adapting the loss function has no clear effect, but globally trained models benefit if the error at each grid-cell is weighted by the respective grid-cell area. A weighted loss function for a globally trained model could also potentially be used to fine-tune the model to specific regions since the weights can also be changed during model training. 
	
	\begin{table}
		\centering
		\caption{RPSS of the weighted models with respect to the unweighted models. The grey shading highlights cases where the weighting yields better models. }
		\centering
		\begin{tabular}{ll|rrrr|rrrr}
			\toprule
			&  & \multicolumn{4}{c}{temperature} & \multicolumn{4}{c}{precipitation} \\
			&  &  global &     NH & tropics &     SH& global &     NH & tropics &     SH \\
			lead time & model &        &        &         &        &        &        &         &        \\
			\midrule
			14 days & UNet global &  \cellcolor{black!15}0.005 &  \cellcolor{black!15}0.001 &   \cellcolor{black!15}0.011 &  \cellcolor{black!15}0.015 &  \cellcolor{black!15}0.001 & 0.000 &   \cellcolor{black!15}0.005 & -0.001 \\ 
			& UNet pw & 0.000 & -0.002 &   \cellcolor{black!15}0.004 & -0.001 & -0.001 & -0.002 &   \cellcolor{black!15}0.002 & -0.002 \\
			& BF-CNN pw &  \cellcolor{black!15}0.004 &  \cellcolor{black!15}0.003 &   \cellcolor{black!15}0.007 & \cellcolor{black!15} 0.012 & 0.000 & -0.002 &   \cellcolor{black!15}0.005 &  \cellcolor{black!15}0.002 \\
			& TC-CNN pw &  0.000 & 0.000 &   \cellcolor{black!15}0.001 &  \cellcolor{black!15}0.002 & -0.001 & -0.003 &   \cellcolor{black!15}0.003 &  \cellcolor{black!15}0.002 \\
			\midrule
			28 days & UNet global &  \cellcolor{black!15}0.005 &  \cellcolor{black!15}0.002 &  \cellcolor{black!15}0.011 &  \cellcolor{black!15}0.015 &  \cellcolor{black!15}0.002 & 0.000 &   \cellcolor{black!15}0.007 & -0.007 \\
			& UNet pw &  \cellcolor{black!15}0.002 &  \cellcolor{black!15}0.001 &   \cellcolor{black!15}0.006 & -0.007 &  0.000 & -0.001 &   \cellcolor{black!15}0.005 & 0.000 \\
			& BF-CNN pw &  \cellcolor{black!15}0.001 &  \cellcolor{black!15}0.001 &   \cellcolor{black!15}0.005 &  \cellcolor{black!15}0.001 & -0.001 & -0.003 &   \cellcolor{black!15}0.004 &  \cellcolor{black!15}0.004 \\
			& TC-CNN pw &  0.000 &  0.000 &   \cellcolor{black!15}0.005 & -0.006 & 0.000 & -0.003 &   \cellcolor{black!15}0.005 &  \cellcolor{black!15}0.004 \\
			\bottomrule
		\end{tabular}
		\label{tab:weighted}
	\end{table}

	\subsection{Critical perspective on tercile approach}\label{criticalTerciles}
	
\begin{figure}[p]
		\centering
		\includegraphics[width=0.82\textwidth,trim={0 0 0 0cm},clip]{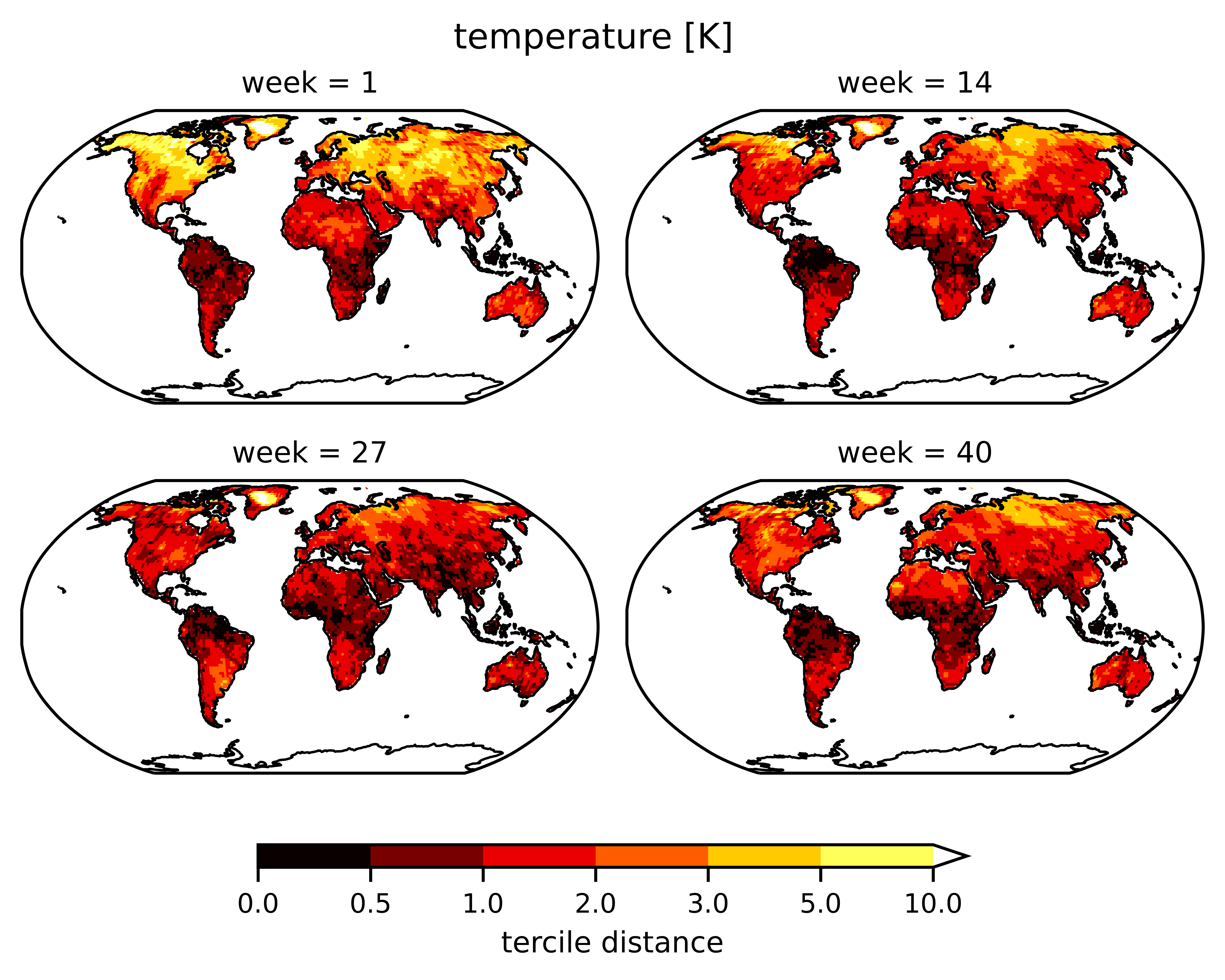}
		
		\includegraphics[width=0.82\textwidth,trim={0 0 0 0cm},clip]{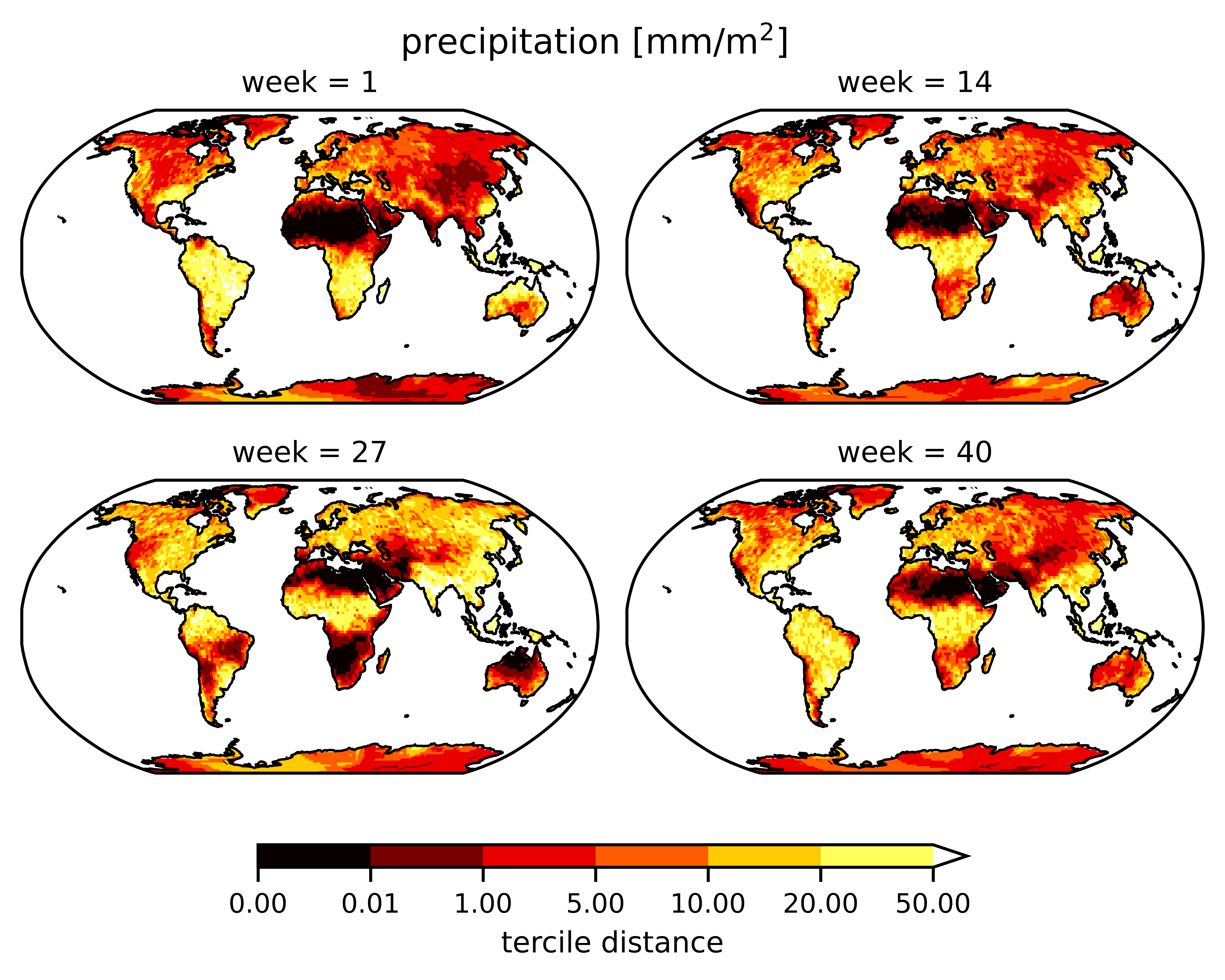}
		\caption{Difference between the upper tercile edge (2/3-quantile) and the lower tercile edge (1/3-quantile) for four selected weeks of the year representing the four seasons. Shown are the tercile edges for mean temperature and accumulated precipitation for weeks 3-4.
		}
		\label{fig:tercile_edge_diff_map}
	\end{figure}

	The example forecasts (Figure \ref{fig:exampleForecasts}) clearly show that the models issue the most prudent forecasts for the middle tercile with probabilities close to climatology. The same behaviour can be observed for the precipitation forecasts (shown in Appendix \ref{appendix:plots}) and agrees with the general understanding that skill for the middle tercile is limited \citep{dool_why_1991, becker_probabilistic_2016}. \citet{wilks_calibration_2013} observed the same behaviour for temperature forecasts over the U.S. Indeed, predicting the middle category correctly is challenging, since the lower and upper tercile edge confining the middle category can be very close together as shown in Figure \ref{fig:tercile_edge_diff_map}. For example in the tropics, the temperature tercile edges differ by less than \SI{1}{K}. Correctly predicting such small temperature fluctuations three to six weeks ahead seems to be very challenging in light of the generally low signal to noise ratio, and likely provides hardly any added value to the forecast user. For precipitation, forecasts in desert areas with category edge differences below \SI{0.01}{mm/m^2} of precipitation in two weeks are presumably also difficult and of low relevance to the forecast user. 
	
	It seems that framing the forecasting task as a tercile classification problem might not always simplify the forecasting problem compared to predicting the actual values of the target variable, but rather adds an another level of complexity. However, forecast quality appears to be independent of the distance between tercile edges, since the correlation between the RPS of the forecasts and the distance between the tercile edges is close to zero. Nevertheless, the benefit of tercile forecasts for variables with very narrow climatological distribution remains questionable. 
	
	\citet{White2017} point out that users of probabilistic tercile forecasts also have to know the usual climatology, i.e., the specific values for the tercile edges to make use of the forecasts, and that such forecasts do not provide actionable information on timing, location and scale of weather events. Further, \citet{White2017} and \citet{brunet_collaboration_2010} argue that S2S predictions should realistically represent the weather and its statistics. The ECMWF baseline forecasts well represent the day-to-day weather conditions, as seen by the fine detailed structure in the example forecast in Figure \ref{fig:exampleForecasts}, but are not calibrated according to Figure \ref{fig:calibration_simplex} and the discussion in Section \ref{calibration}. On the other hand, the post-processed predictions are very smooth and certainly do not represent day-to-day weather, but they are well calibrated. Given the limited predictability on S2S time scales, probabilistic tercile forecasts can presumably not accomplish both, accurately representing the weather statistics, and at the same time provide calibrated probabilities. 
	
	\section{Conclusions}\label{sec:conclusions}
	
	We propose CNN-based post-processing methods to generate spatially homogeneous probabilistic predictions for precipitation and temperature on sub-seasonal time scales. Two of the four presented post-processing methods extend earlier work from \citet{scheuererBasisfunc} to a global scale, and build on the idea of combining CNN models and basis function approaches from geostatistics. The remaining two models adapt the well-known UNet architecture \citep{unet} to the S2S post-processing application. The post-processing models are either trained on collections of small quadratic domains of the global input fields, so-called patches, or directly on the global predictor fields. Spatial fields of ECMWF ensemble mean forecasts of selected meteorological variables thereby serve as input variables (geopotential height at \SI{500}{hPa} and \SI{850}{hPa}, and mean sea-level pressure for temperature, and additionally total column water for precipitation). The post-processing models yield global probabilistic tercile forecasts for biweekly aggregates of temperature and precipitation on a 1.5$^\circ$ grid as output. The setting of our study is based on the S2S AI Challenge \citep{s2saiChallenge} in which our post-processing models (retrospectively) place among the top three submissions. 
	
	All our post-processing models consistently achieve better skill than the calibrated ECMWF baseline forecast and outperform the climatological forecasts for both variables and forecast horizons. The global UNet model performs best and achieves an RPSS of 0.064 for temperature for weeks 3-4 (and of 0.026 for weeks 5-6), and RPSS values of 0.025 and 0.01 for precipitation, respectively. All models including the ECMWF baseline achieve the  highest skill in the tropics (30$^{\circ}$N-30$^{\circ}$S). The post-processed forecasts are substantially less sharp and much smoother than the ECMWF baseline. However, in contrast to the ECMWF forecasts, they are well calibrated and their reduced sharpness appropriately reflects the limited predictability on S2S time scales. The improvements in calibration further demonstrate that our post-processing models enable a reliable quantification of forecast uncertainty based on deterministic input information in the form of the ECMWF ensemble mean forecast fields only.
	The main methodological innovation of our approach is the direct use of spatial input fields without applying any per-grid cell post-processing models. Therefore, our CNN-based methods are able to retain spatial relationships in the forecasts and have the potential to learn spatial error structures, which may explain our particularly good results for precipitation forecasts. With the spatially weighted loss function we employ, the models can in principle be fine-tuned to every region. 
	
	Of the four post-processing approaches we proposed in this study, the globally trained UNet is superior to the others in several ways.
	Firstly, the globally trained UNet is computationally the least expensive. The training of the patch-wise models takes more than seven times longer due to the large number of patches, vastly increasing the number of training samples. We expect the patch-wise training to be beneficial in terms of forecast quality for forecasts with a higher signal to noise ratio as indicated by the slight improvement for the temperature forecasts for weeks 3-4. 
	For forecasts with low signal to noise ratio, the additional training samples do not improve forecast skill of the post-processed forecasts.
	Secondly, the UNet architecture is a standard architecture that can be readily implemented. In particular the global UNet can be easily adapted to other data sets, since no sophisticated data loader is needed.
	These factors make the the global UNet the preferable method for all target variables and lead times in terms of both performance as well as computational costs. 
	
	The proposed post-processing models provide several avenues for future work. An important pathway towards further improvements is the incorporation of additional predictor information. On the one hand, additional spatial fields of meteorological variables such as soil moisture and sea-surface temperature (SST) are known to carry predictability on S2S time scales \citep{White2017, Merryfield2020s2s, robertson2020}. However, including those variables did not improve the predictive performance and comes with additional technical difficulties since for example the SST inputs are only available over the ocean (and soil moisture only over land), which would require the CNN models to learn different filters for land and ocean grid cells. On the other hand, a promising option to generalize and improve our post-processing models is the incorporation of information from slowly varying components of the climate system (e.g., the Madden-Julian Oscillation or the state of the stratosphere) which might enable skillful forecasts on longer time scales via teleconnection pathways \citep{LangEtAl2020}. First tests indicated that combining SST inputs with teleconnection information could lead to further improvements of the post-processing models, however, adaptations of the model architectures will be required to appropriately combine the different types of input information.
	
	Despite the clear improvements over the physical and climatological reference forecasts, the skill of the post-processed forecasts remains limited, with RPSS values on the order of a few percent. To allow for a better assessment of the utility of post-processing on S2S time scales, it would be interesting to further investigate the effects of post-processing on downstream applications of weather forecasts such as fully integrated renewable energy forecasting systems \citep[e.g.,][]{HauptEtAl2020}. 
	For example, in the context of medium-range wind power prediction, it has been demonstrated that post-processing wind speed forecasts can even be detrimental to the quality of the resulting wind power forecasts, calling for a targeted development of application-specific post-processing methods \citep{PhippsEtAl2022}.
	
	
	\section*{Acknowledgments}
	
	The research leading to these results has been done within the Young Investigator Group ``Artificial Intelligence for Probabilistic Weather Forecasting'' funded by the Vector Stiftung. We thank Frederic Vitart for providing Fortran code for the ECMWF baseline, and Johannes Resin for advice on the calibration simplices. We further thank Peter Knippertz, Christian Grams, Julian Quinting, Jieyu Chen, Michael Scheuerer, and all team members contributing to the original submission to the S2S AI Challenge for helpful discussions.
	
	\bibliographystyle{myims2}
	\bibliography{referencesZotero_editedSL, referencesSoftware}

\begin{thebibliography}{55}
\expandafter\ifx\csname natexlab\endcsname\relax\def\natexlab#1{#1}\fi
\expandafter\ifx\csname url\endcsname\relax
  \def\url#1{\texttt{#1}}\fi
\expandafter\ifx\csname urlprefix\endcsname\relax\def\urlprefix{URL }\fi
\providecommand{\eprint}[2][]{\url{#2}}

\bibitem[{Abadi et~al.(2016)Abadi, Barham, Chen, Chen, Davis, Dean, Devin,
  Ghemawat, Irving, Isard et~al.}]{abadi2016tensorflow}
{Abadi, M.}, {Barham, P.}, {Chen, J.}, {Chen, Z.}, {Davis, A.}, {Dean, J.},
  {Devin, M.}, {Ghemawat, S.}, {Irving, G.}, {Isard, M.} {et~al.} (2016).
\newblock Tensorflow: A system for large-scale machine learning.
\newblock In \textit{Proc. USENIX 12th Symposium on Operating Systems Design
  and Implementation}. 265--283.

\bibitem[{Ayzel et~al.(2020)Ayzel, Scheffer and Heistermann}]{ayzel2020}
{Ayzel, G.}, {Scheffer, T.} and {Heistermann, M.} (2020).
\newblock {RainNet} v1.0: A~convolutional neural network for radar-based
  precipitation nowcasting.
\newblock \textit{Geoscientific Model Development}, {13}, 2631--2644.

\bibitem[{Becker and Dool(2016)}]{becker_probabilistic_2016}
{Becker, E.} and {Dool, H. v.~d.} (2016).
\newblock Probabilistic seasonal forecasts in the {North} {American} multimodel
  ensemble: A baseline skill assessment.
\newblock \textit{Journal of Climate}, {29}, 3015--3026.

\bibitem[{Ben-Bouallegue et~al.(2023)Ben-Bouallegue, Weyn, Clare, Dramsch,
  Dueben and Chantry}]{PoET2023}
{Ben-Bouallegue, Z.}, {Weyn, J.~A.}, {Clare, M.~C.}, {Dramsch, J.}, {Dueben,
  P.} and {Chantry, M.} (2023).
\newblock Improving medium-range ensemble weather forecasts with hierarchical
  ensemble transformers.
\newblock \urlprefix\url{https://arxiv.org/abs/2303.17195}.

\bibitem[{Brunet et~al.(2010)Brunet, Shapiro, Hoskins, Moncrieff, Dole,
  Kiladis, Kirtman, Lorenc, Mills, Morss, Polavarapu, Rogers, Schaake and
  Shukla}]{brunet_collaboration_2010}
{Brunet, G.}, {Shapiro, M.}, {Hoskins, B.}, {Moncrieff, M.}, {Dole, R.},
  {Kiladis, G.~N.}, {Kirtman, B.}, {Lorenc, A.}, {Mills, B.}, {Morss, R.},
  {Polavarapu, S.}, {Rogers, D.}, {Schaake, J.} and {Shukla, J.} (2010).
\newblock Collaboration of the weather and climate communities to advance
  subseasonal-to-seasonal prediction.
\newblock \textit{Bulletin of the American Meteorological Society}, {91},
  1397--1406.

\bibitem[{Chapman et~al.(2022)Chapman, Monache, Alessandrini, Subramanian,
  Ralph, Xie, Lerch and Hayatbini}]{Chapman2022}
{Chapman, W.~E.}, {Monache, L.~D.}, {Alessandrini, S.}, {Subramanian, A.~C.},
  {Ralph, F.~M.}, {Xie, S.-P.}, {Lerch, S.} and {Hayatbini, N.} (2022).
\newblock Probabilistic predictions from deterministic atmospheric river
  forecasts with deep learning.
\newblock \textit{Monthly Weather Review}, {150}, 215--234.

\bibitem[{Chollet et~al.(2015)}]{chollet2015keras}
{Chollet, F.} {et~al.} (2015).
\newblock Keras.
\newblock \url{https://keras.io}.
\newblock \urlprefix\url{https://github.com/fchollet/keras}.

\bibitem[{Cireşan et~al.(2012)Cireşan, Giusti, Gambardella and
  Schmidhuber}]{ciresan2012}
{Cireşan, D.~C.}, {Giusti, A.}, {Gambardella, L.~M.} and {Schmidhuber, J.}
  (2012).
\newblock Deep neural networks segment neuronal membranes in electron
  microscopy images.
\newblock In \textit{Advances in {Neural} {Information} {Processing}
  {Systems}}, vol.~4. 2843--2851.

\bibitem[{Dai and Hemri(2021)}]{DaiHemri2021}
{Dai, Y.} and {Hemri, S.} (2021).
\newblock Spatially coherent postprocessing of cloud cover ensemble forecasts.
\newblock \textit{Monthly Weather Review}, {149}, 3923--3937.

\bibitem[{Demaeyer et~al.(2023)Demaeyer, jonas Bhend, Lerch, Primo,
  Schaeybroeck, Atencia, Bouall{\`{e}}gue, Chen, Dabernig, Evans, Pucer,
  Hooper, Horat, Jobst, Mer{\v{s}}e, Mlakar, M\"{o}ller, Mestre, Taillardat and
  Vannitsem}]{DemaeyerEtAl2023}
{Demaeyer, J.}, {jonas Bhend}, {Lerch, S.}, {Primo, C.}, {Schaeybroeck, B.~V.},
  {Atencia, A.}, {Bouall{\`{e}}gue, Z.~B.}, {Chen, J.}, {Dabernig, M.}, {Evans,
  G.}, {Pucer, J.~F.}, {Hooper, B.}, {Horat, N.}, {Jobst, D.}, {Mer{\v{s}}e,
  J.}, {Mlakar, P.}, {M\"{o}ller, A.}, {Mestre, O.}, {Taillardat, M.} and
  {Vannitsem, S.} (2023).
\newblock The {EUPPBench} postprocessing benchmark dataset v1.0.
\newblock \textit{Earth System Science Data}, {15}, in press.

\bibitem[{Doblas-Reyes et~al.(2005)Doblas-Reyes, Hagedorn and
  Palmer}]{doblas-reyes_rationale_2005}
{Doblas-Reyes, F.~J.}, {Hagedorn, R.} and {Palmer, T.~N.} (2005).
\newblock The rationale behind the success of multi-model ensembles in seasonal
  forecasting – {II}. {Calibration} and combination.
\newblock \textit{Tellus A: Dynamic Meteorology and Oceanography}, {57},
  234--252.

\bibitem[{Dool and Toth(1991)}]{dool_why_1991}
{Dool, H. M. V.~D.} and {Toth, Z.} (1991).
\newblock Why do forecasts for “{Near} {Normal}” often fail?
\newblock \textit{Weather and Forecasting}, {6}, 76 -- 85.

\bibitem[{Epstein(1969)}]{rps}
{Epstein, E.~S.} (1969).
\newblock A scoring system for probability forecasts of ranked categories.
\newblock \textit{Journal of Applied Meteorology}, {8}, 985--987.

\bibitem[{Gneiting et~al.(2007)Gneiting, Balabdaoui and
  Raftery}]{GneitingEtAl2007}
{Gneiting, T.}, {Balabdaoui, F.} and {Raftery, A.~E.} (2007).
\newblock Probabilistic forecasts, calibration and sharpness.
\newblock \textit{Journal of the Royal Statistical Society: Series B
  (Statistical Methodology)}, {69}, 243--268.

\bibitem[{Grönquist et~al.(2021)Grönquist, Yao, Ben-Nun, Dryden, Dueben, Li
  and Hoefler}]{gronquist2021}
{Grönquist, P.}, {Yao, C.}, {Ben-Nun, T.}, {Dryden, N.}, {Dueben, P.}, {Li,
  S.} and {Hoefler, T.} (2021).
\newblock Deep learning for post-processing ensemble weather forecasts.
\newblock \textit{Philosophical Transactions of the Royal Society A:
  Mathematical, Physical and Engineering Sciences}, {379}, 20200092.

\bibitem[{Haupt et~al.(2021)Haupt, Chapman, Adams, Kirkwood, Hosking, Robinson,
  Lerch and Subramanian}]{haupt2021}
{Haupt, S.~E.}, {Chapman, W.}, {Adams, S.~V.}, {Kirkwood, C.}, {Hosking,
  J.~S.}, {Robinson, N.~H.}, {Lerch, S.} and {Subramanian, A.~C.} (2021).
\newblock Towards implementing artificial intelligence post-processing in
  weather and climate: Proposed actions from the {Oxford} 2019 workshop.
\newblock \textit{Philosophical Transactions of the Royal Society A:
  Mathematical, Physical and Engineering Sciences}, {379}, 20200091.

\bibitem[{Haupt et~al.(2020)Haupt, McCandless, Dettling, Alessandrini, Lee,
  Linden, Petzke, Brummet, Nguyen and Kosovi{\'c}}]{HauptEtAl2020}
{Haupt, S.~E.}, {McCandless, T.~C.}, {Dettling, S.}, {Alessandrini, S.}, {Lee,
  J.~A.}, {Linden, S.}, {Petzke, W.}, {Brummet, T.}, {Nguyen, N.} and
  {Kosovi{\'c}, B.} (2020).
\newblock Combining artificial intelligence with physics-based methods for
  probabilistic renewable energy forecasting.
\newblock \textit{Energies}, {13}, 1979.

\bibitem[{Hu et~al.(2023)Hu, Ghazvinian, Chapman, Sengupta, Ralph and
  Delle~Monache}]{Hu2023precip}
{Hu, W.}, {Ghazvinian, M.}, {Chapman, W.~E.}, {Sengupta, A.}, {Ralph, F.~M.}
  and {Delle~Monache, L.} (2023).
\newblock Deep learning forecast uncertainty for precipitation over {Western}
  {US}.
\newblock \textit{Monthly Weather Review}, {151}, 1367--1385.

\bibitem[{Kingma and Ba(2017)}]{adam2015}
{Kingma, D.~P.} and {Ba, J.} (2017).
\newblock Adam: A method for stochastic optimization.
\newblock \urlprefix\url{https://arxiv.org/abs/1412.6980}.

\bibitem[{Lagerquist et~al.(2021)Lagerquist, Stewart, Ebert-Uphoff and
  Kumler}]{lagerquist2021}
{Lagerquist, R.}, {Stewart, J.~Q.}, {Ebert-Uphoff, I.} and {Kumler, C.} (2021).
\newblock Using deep learning to nowcast the spatial coverage of convection
  from {Himawari-8} satellite data.
\newblock \textit{Monthly Weather Review}, {149}, 3897 -- 3921.

\bibitem[{Lang et~al.(2020)Lang, Pegion and Barnes}]{LangEtAl2020}
{Lang, A.~L.}, {Pegion, K.} and {Barnes, E.~A.} (2020).
\newblock {Introduction to special collection: ``Bridging weather and climate:
  Subseasonal-to-seasonal (S2S) prediction''}.
\newblock \textit{Journal of Geophysical Research: Atmospheres}, {125},
  e2019JD031833.

\bibitem[{Lerch and Polsterer(2022)}]{LerchPolsterer2022}
{Lerch, S.} and {Polsterer, K.~L.} (2022).
\newblock Convolutional autoencoders for spatially-informed ensemble
  post-processing.
\newblock International Conference on Learning Representations (ICLR) 2022 - AI
  for Earth and Space Science Workshop,
  \urlprefix\url{https://arxiv.org/abs/2204.05102}.

\bibitem[{Li et~al.(2022)Li, Pan, Xia and Duan}]{Li2022precip}
{Li, W.}, {Pan, B.}, {Xia, J.} and {Duan, Q.} (2022).
\newblock Convolutional neural network-based statistical post-processing of
  ensemble precipitation forecasts.
\newblock \textit{Journal of Hydrology}, {605}, 127301.

\bibitem[{Li et~al.(2021)Li, Tian and Medina}]{li2017MME}
{Li, Y.}, {Tian, D.} and {Medina, H.} (2021).
\newblock Multimodel subseasonal precipitation forecasts over the contiguous
  {United} {States}: Skill assessment and statistical postprocessing.
\newblock \textit{Journal of Hydrometeorology}, {22}, 2581–2600.

\bibitem[{Litjens et~al.(2017)Litjens, Kooi, Bejnordi, Setio, Ciompi,
  Ghafoorian, van~der Laak, van Ginneken and Sánchez}]{imageSegmetationReview}
{Litjens, G.}, {Kooi, T.}, {Bejnordi, B.~E.}, {Setio, A. A.~A.}, {Ciompi, F.},
  {Ghafoorian, M.}, {van~der Laak, J.~A.}, {van Ginneken, B.} and {Sánchez,
  C.~I.} (2017).
\newblock A survey on deep learning in medical image analysis.
\newblock \textit{Medical Image Analysis}, {42}, 60--88.

\bibitem[{Liu et~al.(2018)Liu, Ren, Geng, Ding and Li}]{patchwiseTraining}
{Liu, Y.}, {Ren, Q.}, {Geng, J.}, {Ding, M.} and {Li, J.} (2018).
\newblock Efficient patch-wise semantic segmentation for large-scale remote
  sensing images.
\newblock \textit{Sensors}, {18}, 3232.

\bibitem[{Mariotti et~al.(2020)Mariotti, Baggett, Barnes, Becker, Butler,
  Collins, Dirmeyer, Ferranti, Johnson, Jones, Kirtman, Lang, Molod, Newman,
  Robertson, Schubert, Waliser and Albers}]{Mariotti2020}
{Mariotti, A.}, {Baggett, C.}, {Barnes, E.~A.}, {Becker, E.}, {Butler, A.},
  {Collins, D.~C.}, {Dirmeyer, P.~A.}, {Ferranti, L.}, {Johnson, N.~C.},
  {Jones, J.}, {Kirtman, B.~P.}, {Lang, A.~L.}, {Molod, A.}, {Newman, M.},
  {Robertson, A.~W.}, {Schubert, S.}, {Waliser, D.~E.} and {Albers, J.} (2020).
\newblock Windows of opportunity for skillful forecasts subseasonal to seasonal
  and beyond.
\newblock \textit{Bulletin of the American Meteorological Society}, {101},
  E608--E625.

\bibitem[{Mayer and Barnes(2021)}]{mayer_subseasonal_2021}
{Mayer, K.~J.} and {Barnes, E.~A.} (2021).
\newblock Subseasonal forecasts of opportunity identified by an explainable
  neural network.
\newblock \textit{Geophysical Research Letters}, {48}, e2020GL092092.

\bibitem[{Merryfield et~al.(2020)Merryfield, Baehr, Batté, Becker, Butler,
  Coelho, Danabasoglu, Dirmeyer, Doblas-Reyes, Domeisen, Ferranti, Ilynia,
  Kumar, Müller, Rixen, Robertson, Smith, Takaya, Tuma, Vitart, White,
  Alvarez, Ardilouze, Attard, Baggett, Balmaseda, Beraki, Bhattacharjee,
  Bilbao, de~Andrade, DeFlorio, Díaz, Ehsan, Fragkoulidis, Grainger, Green,
  Hell, Infanti, Isensee, Kataoka, Kirtman, Klingaman, Lee, Mayer, McKay,
  Mecking, Miller, Neddermann, Justin~Ng, Ossó, Pankatz, Peatman, Pegion,
  Perlwitz, Recalde-Coronel, Reintges, Renkl, Solaraju-Murali, Spring, Stan,
  Sun, Tozer, Vigaud, Woolnough and Yeager}]{Merryfield2020s2s}
{Merryfield, W.~J.}, {Baehr, J.}, {Batté, L.}, {Becker, E.~J.}, {Butler,
  A.~H.}, {Coelho, C. A.~S.}, {Danabasoglu, G.}, {Dirmeyer, P.~A.},
  {Doblas-Reyes, F.~J.}, {Domeisen, D. I.~V.}, {Ferranti, L.}, {Ilynia, T.},
  {Kumar, A.}, {Müller, W.~A.}, {Rixen, M.}, {Robertson, A.~W.}, {Smith,
  D.~M.}, {Takaya, Y.}, {Tuma, M.}, {Vitart, F.}, {White, C.~J.}, {Alvarez,
  M.~S.}, {Ardilouze, C.}, {Attard, H.}, {Baggett, C.}, {Balmaseda, M.~A.},
  {Beraki, A.~F.}, {Bhattacharjee, P.~S.}, {Bilbao, R.}, {de~Andrade, F.~M.},
  {DeFlorio, M.~J.}, {Díaz, L.~B.}, {Ehsan, M.~A.}, {Fragkoulidis, G.},
  {Grainger, S.}, {Green, B.~W.}, {Hell, M.~C.}, {Infanti, J.~M.}, {Isensee,
  K.}, {Kataoka, T.}, {Kirtman, B.~P.}, {Klingaman, N.~P.}, {Lee, J.-Y.},
  {Mayer, K.}, {McKay, R.}, {Mecking, J.~V.}, {Miller, D.~E.}, {Neddermann,
  N.}, {Justin~Ng, C.~H.}, {Ossó, A.}, {Pankatz, K.}, {Peatman, S.}, {Pegion,
  K.}, {Perlwitz, J.}, {Recalde-Coronel, G.~C.}, {Reintges, A.}, {Renkl, C.},
  {Solaraju-Murali, B.}, {Spring, A.}, {Stan, C.}, {Sun, Y.~Q.}, {Tozer,
  C.~R.}, {Vigaud, N.}, {Woolnough, S.} and {Yeager, S.} (2020).
\newblock Current and emerging developments in subseasonal to decadal
  prediction.
\newblock \textit{Bulletin of the American Meteorological Society}, {101},
  E869--E896.

\bibitem[{Messner et~al.(2017)Messner, Mayr and Zeileis}]{MessnerEtAl2017}
{Messner, J.~W.}, {Mayr, G.~J.} and {Zeileis, A.} (2017).
\newblock Nonhomogeneous boosting for predictor selection in ensemble
  postprocessing.
\newblock \textit{Monthly Weather Review}, {145}, 137--147.

\bibitem[{Mouatadid et~al.(2021)Mouatadid, Orenstein, Flaspohler, Oprescu,
  Cohen, Wang, Knight, Geogdzhayeva, Levang, Fraenkel and
  Mackey}]{mouatadid_learned_2021}
{Mouatadid, S.}, {Orenstein, P.}, {Flaspohler, G.}, {Oprescu, M.}, {Cohen, J.},
  {Wang, F.}, {Knight, S.}, {Geogdzhayeva, M.}, {Levang, S.}, {Fraenkel, E.}
  and {Mackey, L.} (2021).
\newblock Learned benchmarks for subseasonal forecasting.
\newblock \urlprefix\url{https://arxiv.org/abs/2109.10399}.

\bibitem[{Murphy(1971)}]{rpsMurphy}
{Murphy, A.~H.} (1971).
\newblock A note on the ranked probability score.
\newblock \textit{Journal of Applied Meteorology}, {10}, 155--156.

\bibitem[{Murphy and Winkler(1977)}]{murphy_reliability_1977}
{Murphy, A.~H.} and {Winkler, R.~L.} (1977).
\newblock Reliability of subjective probability forecasts of precipitation and
  temperature.
\newblock \textit{Applied Statistics}, {26}, 41.

\bibitem[{Phipps et~al.(2022)Phipps, Lerch, Andersson, Mikut, Hagenmeyer and
  Ludwig}]{PhippsEtAl2022}
{Phipps, K.}, {Lerch, S.}, {Andersson, M.}, {Mikut, R.}, {Hagenmeyer, V.} and
  {Ludwig, N.} (2022).
\newblock Evaluating ensemble post-processing for wind power forecasts.
\newblock \textit{Wind Energy}, {25}, 1379--1405.

\bibitem[{Quinting and Grams(2022)}]{quinting2022}
{Quinting, J.~F.} and {Grams, C.~M.} (2022).
\newblock {EuLerian} {Identification} of ascending {AirStreams} ({ELIAS} 2.0)
  in numerical weather prediction and climate models -- part 1: Development of
  deep learning model.
\newblock \textit{Geoscientific Model Development}, {15}, 715--730.

\bibitem[{Rasp and Lerch(2018)}]{rasp_neural_2018}
{Rasp, S.} and {Lerch, S.} (2018).
\newblock Neural networks for postprocessing ensemble weather forecasts.
\newblock \textit{Monthly Weather Review}, {146}, 3885--3900.

\bibitem[{Resin(2021)}]{resin2021calsim}
{Resin, J.} (2021).
\newblock \textit{CalSim: The Calibration Simplex}.
\newblock R package version 0.5.2.

\bibitem[{Robertson et~al.(2020)Robertson, Vitart and Camargo}]{robertson2020}
{Robertson, A.~W.}, {Vitart, F.} and {Camargo, S.~J.} (2020).
\newblock Subseasonal to seasonal prediction of weather to climate with
  application to tropical cyclones.
\newblock \textit{Journal of Geophysical Research: Atmospheres}, {125},
  e2018JD029375.

\bibitem[{Ronneberger et~al.(2015)Ronneberger, Fischer and Brox}]{unet}
{Ronneberger, O.}, {Fischer, P.} and {Brox, T.} (2015).
\newblock U-{Net}: Convolutional networks for biomedical image segmentation.
\newblock In \textit{Medical {Image} {Computing} and {Computer}-{Assisted}
  {Intervention} – {MICCAI} 2015} (N.~Navab, J.~Hornegger, W.~M. Wells and
  A.~F. Frangi, eds.). Springer International Publishing, Cham, 234--241.

\bibitem[{Sacco et~al.(2022)Sacco, Ruiz, Pulido and
  Tandeo}]{sacco_evaluation_2022}
{Sacco, M.~A.}, {Ruiz, J.~J.}, {Pulido, M.} and {Tandeo, P.} (2022).
\newblock Evaluation of machine learning techniques for forecast uncertainty
  quantification.
\newblock \textit{Quarterly Journal of the Royal Meteorological Society},
  {148}, 3470--3490.

\bibitem[{Scheuerer et~al.(2020)Scheuerer, Switanek, Worsnop and
  Hamill}]{scheuererBasisfunc}
{Scheuerer, M.}, {Switanek, M.~B.}, {Worsnop, R.~P.} and {Hamill, T.~M.}
  (2020).
\newblock Using artificial neural networks for generating probabilistic
  subseasonal precipitation forecasts over {California}.
\newblock \textit{Monthly Weather Review}, {148}, 3489--3506.

\bibitem[{Schulz and Lerch(2022)}]{SchulzLerch2022}
{Schulz, B.} and {Lerch, S.} (2022).
\newblock Machine learning methods for postprocessing ensemble forecasts of
  wind gusts: A systematic comparison.
\newblock \textit{Monthly Weather Review}, {150}, 235--257.

\bibitem[{Taillardat et~al.(2016)Taillardat, Mestre, Zamo and
  Naveau}]{TaillardatEtAl2016}
{Taillardat, M.}, {Mestre, O.}, {Zamo, M.} and {Naveau, P.} (2016).
\newblock Calibrated ensemble forecasts using quantile regression forests and
  ensemble model output statistics.
\newblock \textit{Monthly Weather Review}, {144}, 2375--2393.

\bibitem[{Vannitsem et~al.(2021)Vannitsem, Bremnes, Demaeyer, Evans, Flowerdew,
  Hemri, Lerch, Roberts, Theis, Atencia, Ben~Bouallègue, Bhend, Dabernig,
  De~Cruz, Hieta, Mestre, Moret, Plenković, Schmeits, Taillardat, Van~den
  Bergh, Van~Schaeybroeck, Whan and Ylhaisi}]{vannitsem_statistical_2021}
{Vannitsem, S.}, {Bremnes, J.~B.}, {Demaeyer, J.}, {Evans, G.~R.}, {Flowerdew,
  J.}, {Hemri, S.}, {Lerch, S.}, {Roberts, N.}, {Theis, S.}, {Atencia, A.},
  {Ben~Bouallègue, Z.}, {Bhend, J.}, {Dabernig, M.}, {De~Cruz, L.}, {Hieta,
  L.}, {Mestre, O.}, {Moret, L.}, {Plenković, I.~O.}, {Schmeits, M.},
  {Taillardat, M.}, {Van~den Bergh, J.}, {Van~Schaeybroeck, B.}, {Whan, K.} and
  {Ylhaisi, J.} (2021).
\newblock Statistical postprocessing for weather forecasts: Review, challenges,
  and avenues in a big data world.
\newblock \textit{Bulletin of the American Meteorological Society}, {102},
  E681--E699.

\bibitem[{Veldkamp et~al.(2021)Veldkamp, Whan, Dirksen and
  Schmeits}]{Veldkamp2021wind}
{Veldkamp, S.}, {Whan, K.}, {Dirksen, S.} and {Schmeits, M.} (2021).
\newblock Statistical postprocessing of wind speed forecasts using
  convolutional neural networks.
\newblock \textit{Monthly Weather Review}, {149}, 1141--1152.

\bibitem[{Vigaud et~al.(2017)Vigaud, Robertson and
  Tippett}]{Vigaud2017MMEprecip}
{Vigaud, N.}, {Robertson, A.~W.} and {Tippett, M.~K.} (2017).
\newblock Multimodel ensembling of subseasonal precipitation forecasts over
  {North} {America}.
\newblock \textit{Monthly Weather Review}, {145}, 3913--3928.

\bibitem[{Vigaud et~al.(2019)Vigaud, Tippett, Yuan, Robertson and
  Acharya}]{Vigaud2019MMEtemperature}
{Vigaud, N.}, {Tippett, M.~K.}, {Yuan, J.}, {Robertson, A.~W.} and {Acharya,
  N.} (2019).
\newblock Probabilistic skill of subseasonal surface temperature forecasts over
  {North} {America}.
\newblock \textit{Weather and Forecasting}, {34}, 1789--1806.

\bibitem[{Vitart et~al.(2017)Vitart, Ardilouze, Bonet, Brookshaw, Chen,
  Codorean, Déqué, Ferranti, Fucile, Fuentes, Hendon, Hodgson, Kang, Kumar,
  Lin, Liu, Liu, Malguzzi, Mallas, Manoussakis, Mastrangelo, MacLachlan,
  McLean, Minami, Mladek, Nakazawa, Najm, Nie, Rixen, Robertson, Ruti, Sun,
  Takaya, Tolstykh, Venuti, Waliser, Woolnough, Wu, Won, Xiao, Zaripov and
  Zhang}]{s2sProject}
{Vitart, F.}, {Ardilouze, C.}, {Bonet, A.}, {Brookshaw, A.}, {Chen, M.},
  {Codorean, C.}, {Déqué, M.}, {Ferranti, L.}, {Fucile, E.}, {Fuentes, M.},
  {Hendon, H.}, {Hodgson, J.}, {Kang, H.-S.}, {Kumar, A.}, {Lin, H.}, {Liu,
  G.}, {Liu, X.}, {Malguzzi, P.}, {Mallas, I.}, {Manoussakis, M.},
  {Mastrangelo, D.}, {MacLachlan, C.}, {McLean, P.}, {Minami, A.}, {Mladek,
  R.}, {Nakazawa, T.}, {Najm, S.}, {Nie, Y.}, {Rixen, M.}, {Robertson, A.~W.},
  {Ruti, P.}, {Sun, C.}, {Takaya, Y.}, {Tolstykh, M.}, {Venuti, F.}, {Waliser,
  D.}, {Woolnough, S.}, {Wu, T.}, {Won, D.-J.}, {Xiao, H.}, {Zaripov, R.} and
  {Zhang, L.} (2017).
\newblock The subseasonal to seasonal ({S2S}) prediction project database.
\newblock \textit{Bulletin of the American Meteorological Society}, {98},
  163--173.

\bibitem[{Vitart et~al.(2022)Vitart, Robertson, Spring, Pinault, Roškar, Cao,
  Bech, Bienkowski, Caltabiano, De~Coning, Denis, Dirkson, Dramsch, Dueben,
  Gierschendorf, Kim, Nowak, Landry, Lledó, Palma, Rasp and
  Zhou}]{s2saiChallenge}
{Vitart, F.}, {Robertson, A.~W.}, {Spring, A.}, {Pinault, F.}, {Roškar, R.},
  {Cao, W.}, {Bech, S.}, {Bienkowski, A.}, {Caltabiano, N.}, {De~Coning, E.},
  {Denis, B.}, {Dirkson, A.}, {Dramsch, J.}, {Dueben, P.}, {Gierschendorf, J.},
  {Kim, H.~S.}, {Nowak, K.}, {Landry, D.}, {Lledó, L.}, {Palma, L.}, {Rasp,
  S.} and {Zhou, S.} (2022).
\newblock Outcomes of the {WMO} prize challenge to improve subseasonal to
  seasonal predictions using artificial intelligence.
\newblock \textit{Bulletin of the American Meteorological Society}, {103},
  E2878--E2886.

\bibitem[{Weyn et~al.(2020)Weyn, Durran and Caruana}]{weyn_improving_2020}
{Weyn, J.~A.}, {Durran, D.~R.} and {Caruana, R.} (2020).
\newblock Improving data‐driven global weather prediction using deep
  convolutional neural networks on a cubed sphere.
\newblock \textit{Journal of Advances in Modeling Earth Systems}, {12},
  e2020MS002109.

\bibitem[{White et~al.(2017)White, Carlsen, Robertson, Klein, Lazo, Kumar,
  Vitart, Coughlan~de Perez, Ray, Murray, Bharwani, MacLeod, James, Fleming,
  Morse, Eggen, Graham, Kjellström, Becker, Pegion, Holbrook, McEvoy,
  Depledge, Perkins-Kirkpatrick, Brown, Street, Jones, Remenyi,
  Hodgson-Johnston, Buontempo, Lamb, Meinke, Arheimer and Zebiak}]{White2017}
{White, C.~J.}, {Carlsen, H.}, {Robertson, A.~W.}, {Klein, R.~J.}, {Lazo,
  J.~K.}, {Kumar, A.}, {Vitart, F.}, {Coughlan~de Perez, E.}, {Ray, A.~J.},
  {Murray, V.}, {Bharwani, S.}, {MacLeod, D.}, {James, R.}, {Fleming, L.},
  {Morse, A.~P.}, {Eggen, B.}, {Graham, R.}, {Kjellström, E.}, {Becker, E.},
  {Pegion, K.~V.}, {Holbrook, N.~J.}, {McEvoy, D.}, {Depledge, M.},
  {Perkins-Kirkpatrick, S.}, {Brown, T.~J.}, {Street, R.}, {Jones, L.},
  {Remenyi, T.~A.}, {Hodgson-Johnston, I.}, {Buontempo, C.}, {Lamb, R.},
  {Meinke, H.}, {Arheimer, B.} and {Zebiak, S.~E.} (2017).
\newblock Potential applications of subseasonal-to-seasonal ({S2S})
  predictions.
\newblock \textit{Meteorological Applications}, {24}, 315--325.

\bibitem[{White et~al.(2022)White, Domeisen, Acharya, Adefisan, Anderson, Aura,
  Balogun, Bertram, Bluhm, Brayshaw, Browell, Büeler, Charlton-Perez, Chourio,
  Christel, Coelho, DeFlorio, Delle~Monache, Di~Giuseppe, García-Solórzano,
  Gibson, Goddard, González~Romero, Graham, Graham, Grams, Halford, Huang,
  Jensen, Kilavi, Lawal, Lee, MacLeod, Manrique-Suñén, Martins, Maxwell,
  Merryfield, Muñoz, Olaniyan, Otieno, Oyedepo, Palma, Pechlivanidis, Pons,
  Ralph, Reis, Remenyi, Risbey, Robertson, Robertson, Smith, Soret, Sun, Todd,
  Tozer, Vasconcelos, Vigo, Waliser, Wetterhall and
  Wilson}]{White2022applications}
{White, C.~J.}, {Domeisen, D. I.~V.}, {Acharya, N.}, {Adefisan, E.~A.},
  {Anderson, M.~L.}, {Aura, S.}, {Balogun, A.~A.}, {Bertram, D.}, {Bluhm, S.},
  {Brayshaw, D.~J.}, {Browell, J.}, {Büeler, D.}, {Charlton-Perez, A.},
  {Chourio, X.}, {Christel, I.}, {Coelho, C. A.~S.}, {DeFlorio, M.~J.},
  {Delle~Monache, L.}, {Di~Giuseppe, F.}, {García-Solórzano, A.~M.}, {Gibson,
  P.~B.}, {Goddard, L.}, {González~Romero, C.}, {Graham, R.~J.}, {Graham,
  R.~M.}, {Grams, C.~M.}, {Halford, A.}, {Huang, W. T.~K.}, {Jensen, K.},
  {Kilavi, M.}, {Lawal, K.~A.}, {Lee, R.~W.}, {MacLeod, D.}, {Manrique-Suñén,
  A.}, {Martins, E. S. P.~R.}, {Maxwell, C.~J.}, {Merryfield, W.~J.}, {Muñoz,
  A.~G.}, {Olaniyan, E.}, {Otieno, G.}, {Oyedepo, J.~A.}, {Palma, L.},
  {Pechlivanidis, I.~G.}, {Pons, D.}, {Ralph, F.~M.}, {Reis, D.~S.}, {Remenyi,
  T.~A.}, {Risbey, J.~S.}, {Robertson, D. J.~C.}, {Robertson, A.~W.}, {Smith,
  S.}, {Soret, A.}, {Sun, T.}, {Todd, M.~C.}, {Tozer, C.~R.}, {Vasconcelos,
  F.~C.}, {Vigo, I.}, {Waliser, D.~E.}, {Wetterhall, F.} and {Wilson, R.~G.}
  (2022).
\newblock Advances in the application and utility of subseasonal-to-seasonal
  predictions.
\newblock \textit{Bulletin of the American Meteorological Society}, {103},
  E1448--E1472.

\bibitem[{Wilks(2013)}]{wilks_calibration_2013}
{Wilks, D.~S.} (2013).
\newblock The {Calibration} {Simplex}: A generalization of the reliability
  diagram for three-category probability forecasts.
\newblock \textit{Weather and Forecasting}, {28}, 1210--1218.

\bibitem[{Wilks(2020)}]{wilks2011}
{Wilks, D.~S.} (2020).
\newblock \textit{Statistical methods in the atmospheric sciences}.
\newblock 4th ed. Elsevier.

\bibitem[{Zhang et~al.(2023)Zhang, Yang, Gao, Hong, Zhang, Wen and
  Cheng}]{zhang_improving_2023}
{Zhang, L.}, {Yang, T.}, {Gao, S.}, {Hong, Y.}, {Zhang, Q.}, {Wen, X.} and
  {Cheng, C.} (2023).
\newblock Improving subseasonal-to-seasonal forecasts in predicting the
  occurrence of extreme precipitation events over the contiguous {U}.{S}. using
  machine learning models.
\newblock \textit{Atmospheric Research}, {281}, 106502.

\end{thebibliography}
	
	\newpage
	\begin{appendices}
		
		\section{Evaluation metrics}\label{evalMethods}
		We evaluate our predictions with the ranked probability skill score (RPSS;  \citealp{rps, rpsMurphy}).  This skill score is based on the ranked probability score (RPS) which is a strictly proper scoring rule for verifying probabilistic multi-category forecasts. It computes the squared difference of the forecast cumulative density function $F_{m}$  and the observation vector $O_{m}$, 
		
		\begin{equation*}
		\text{RPS} = \sum_{m=1}^{M} (F_{m} - O_{m})^2.
		\end{equation*}
		M denotes the number of categories, and is here equal to three (below normal, near normal, above normal). A perfect forecast achieves an RPS of 0. For details, see \cite{wilks2011}.
		
		The RPSS compares the RPS of the ML prediction $f$ with the score obtained by a reference forecast, which is in our case a climatological forecast that assigns equal probabilities of 1/3 to all three outcomes. The RPSS is computed as 
		\begin{equation}
		\text{RPSS}(f) = \frac{\text{RPS}_\text{ref} - \text{RPS}_{f}}{\text{RPS}_\text{ref} - \text{RPS}_\text{opt}},
		\end{equation}
		where $\text{RPS}_\text{opt} = 0$. Skill scores larger than zero correspond to skillful forecasts.  Skill scores are positively oriented, larger values thus indicate better forecasts.
		
		\section{Further details on model training and hyperparameters}\label{appendix:hyperparams}
		
		Model hyperparameters were tuned using cross-validation resulting in different combinations depending on model architecture, target variable and training mode (patch-wise or global). Hyperparameters related to the generation of the training batches are shown in Table \ref{tab:batchGeneration}, parameters related more closely to model training are detailed in Table \ref{tab:trainingModalities}. These hyperparameter choices will be discussed in the following. 
		
		\subsection{Padding} \label{appendix:padding}
		For the global training, we pad the input fields with eight grid-cells on every side to avoid large discontinuities at the date line and the poles. For the patch-wise training, we need the padding of the input fields to ensure that the smaller output patches (i.e., the predictions) cover the whole globe.  We pad the global fields on all sides before the patch selection according to the difference between input and output patch size as well as the output patch size itself,
		\begin{equation*}
		\text{pad} = (ps_\text{input} - ps_\text{output})/2  + ps_\text{output},
		\end{equation*}
		where $ps$ refers to the patch size, i.e., the number of grid cells in latitude and longitude direction each. The corresponding hyperparameters for global and patch-wise training are detailed in Table \ref{tab:batchGeneration}.
		
		\begin{table}
			\caption{Hyperparameters related to the generation of training batches. }
			\label{tab:batchGeneration}
			\begin{center}
				\begin{tabular}{lcccc}
					\toprule
					Parameter & UNet global & UNet pw & BF-CNN pw & TC-CNN pw \\
					\midrule
					Patch size (input) & Global + 16 & 32 & 34 & 34 \\
					Patch size (output) & Global & 24 & 8 & 16 \\
					Padding & 8 & 28 & 21 & 25 \\
					\midrule
					Patch stride & None & 12 & 12 & 12 \\
					Fraction of NA & - & 0.5 & 0.5 & 0.5 \\
					\midrule
					Batch size & 16 & 32 & 64 & 64 \\
					\bottomrule
				\end{tabular}
			\end{center}
		\end{table}
		
		\subsection{Early stopping and learn rate decay}\label{appendix:earlystop}
		
		Using standard early stopping did not work for precipitation since some validation years were particularly hard to predict and training would stop already after the first epoch. Still, we assumed that a model trained for more than one epoch would perform better than a barely trained model, which is why we introduce delayed early stopping. With delayed early stopping, the stopping criteria is checked for the first time at the end of the second epoch for the patch-wise models and at the end of the fifth epoch for the global model (with the epoch count starting at 0). Further, the learning rate was adapted during training to avoid that the models get stuck in local optima. For precipitation, we introduced learning rate decay since this seemed to improve the generalization ability of the models.
		The final configurations are summarized in Table \ref{tab:trainingModalities}.
		
		\begin{table}
			\caption{Hyperparameters related to model training.}
			\label{tab:trainingModalities}
			\begin{center}
				\begin{tabular}{lcccc}
					\toprule
					\multirow{2}{*}{Parameter} & \multicolumn{2}{c}{Global} & \multicolumn{2}{c}{Patch-wise} \\\rule[-3mm]{0mm}{3mm}& Temperature & Precipitation & Temperature &  Precipitation\\
					\midrule
					Epochs & 50 & 50 & 20 & 20  \\Early stopping & True & True & True & True \\
					Patience & 10 & 10 & 3 & 3 \\
					\midrule
					Epoch to start early stopping & 0 & 5 & 0 & 2\\ 
					Learning rate & 1e-4 & 0.001 & 1e-4 & 0.001 \\
					Learning rate decay & 0 & 0.005& 0 & 0.005 \\
					\bottomrule
				\end{tabular}
			\end{center}
		\end{table}
		
		\newpage
		
		\section{Additional Figure}\label{appendix:plots}
		
		\begin{figure}[!htb]
			\centering
			\includegraphics[width=0.98\textwidth,trim={0 4.05cm 0 0cm},clip]{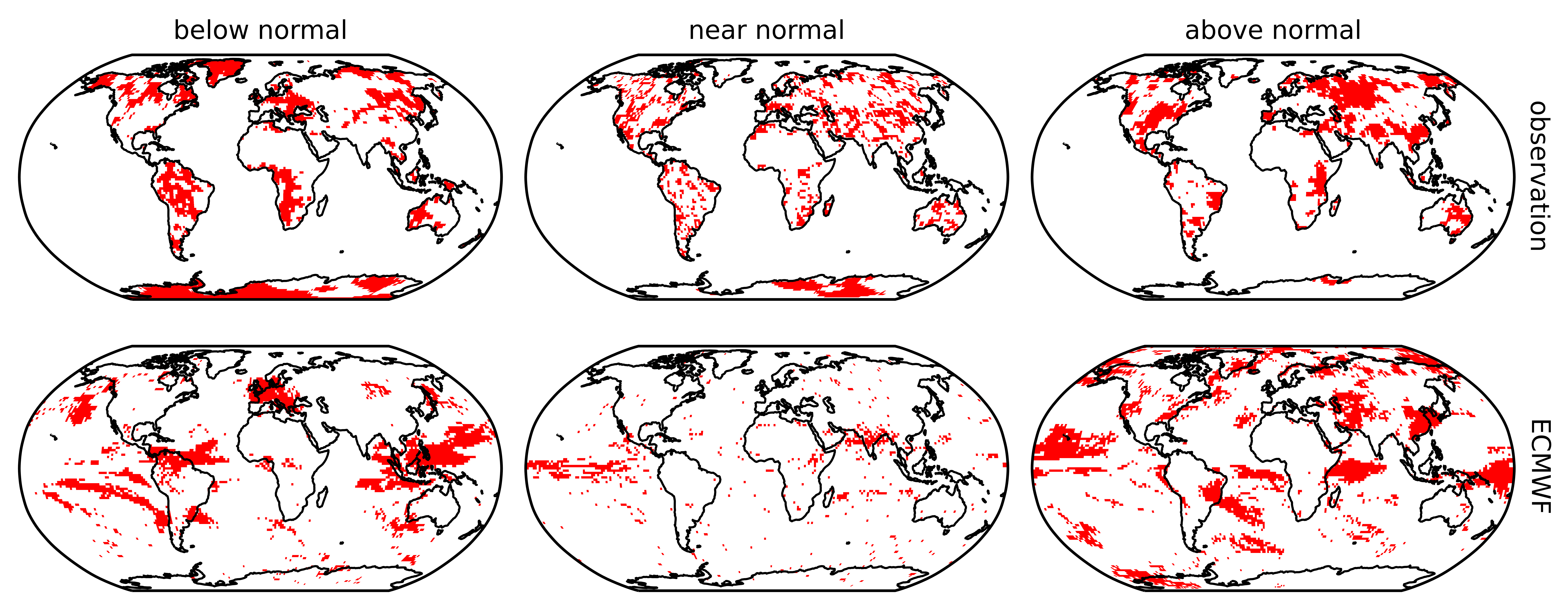}
			\includegraphics[width=0.13\textwidth,trim={11.5cm 7.85cm 0.55cm 1.5cm},clip]{example_obs_terc_legend.png}\hspace*{8cm}
			
			\vspace{0.2cm}
			\includegraphics[width=0.98\textwidth,
			clip]{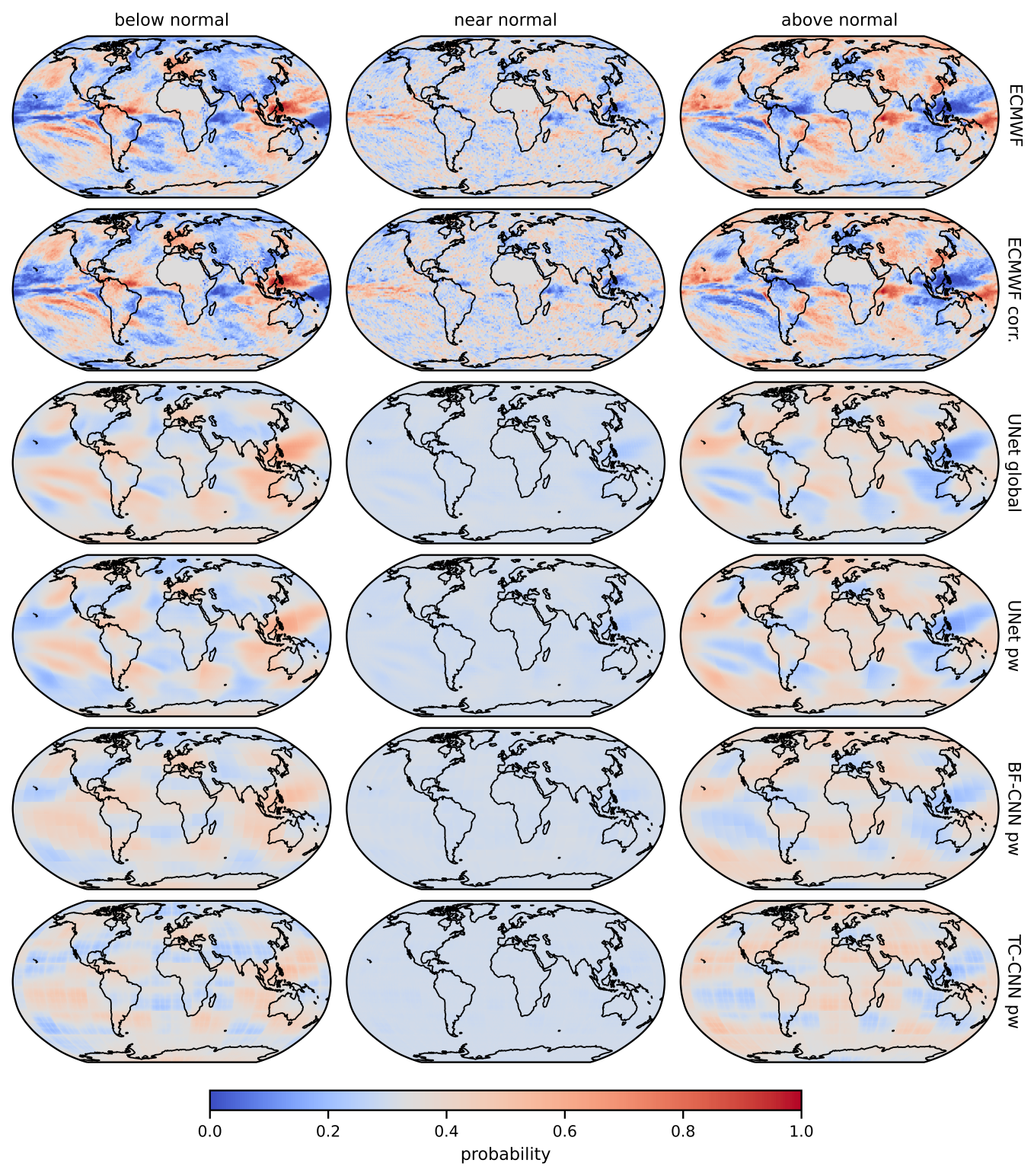}
			\caption{Example predictions for precipitation with a lead time of 14 days issued on 2 January 2020. 
			}
			\label{fig:exampleForecasts_tp_0}
		\end{figure}
		
	\end{appendices}

\end{document}